\documentclass[a4paper,10pt,fleqn]{article}

\usepackage{graphicx}
\usepackage{hyperref}
\usepackage{breakurl}
\usepackage{natbib}
\usepackage{setspace}
\usepackage{amsmath}
\usepackage{amssymb}
\usepackage{mathtools}
\usepackage{datetime}
\usepackage[bottom]{footmisc}
\RequirePackage[english,iso]{isodate}

\newenvironment{myitemize}
{\begin{itemize}
  \setlength{\itemsep}{3.0pt}
  \setlength{\parskip}{0pt}
  \setlength{\parsep}{0pt}}
{\end{itemize}}

\newenvironment{myenumerate}
{\begin{enumerate}
  \setlength{\itemsep}{3.0pt}
  \setlength{\parskip}{0pt}
  \setlength{\parsep}{0pt}}
{\end{enumerate}}

\def\arcsec{arcsec}
\def\parallax{\varpi}
\def\parsd{\sigma_\varpi}
\def\fpe{{\mathnormal f}}
\def\fpetrue{\fpe_{\rm true}}
\def\rtrue{r_{\rm true}}
\def\rest{r_{\!\rm est}}
\def\rmin{r_{\rm min}}
\def\rmode{r_{\rm mode}}

\def\rmed{r_{\rm med}}
\def\rlim{r_{\rm lim}}  
\def\rlen{L} 
\def\like{P(\parallax \,|\, r, \parsd)}
\newcommand{\uprior}[1][]{P^*_{\rm #1}(r)}
\newcommand{\prior}[1][]{P_{\rm #1}(r)}
\newcommand{\upost}[1][]{P^*_{\rm #1}(r \,|\, \parallax, \parsd)}
\newcommand{\post}[1][]{P_{\rm #1}(r \,|\, \parallax, \parsd)}

\begin{document}

\vspace*{1cm}

{\LARGE {\bf Estimating distances from parallaxes}}

\vspace*{0.5cm}
Coryn A.L.\ Bailer-Jones\\
Max Planck Institute for Astronomy, Heidelberg\\
\url{http://www.mpia.de/homes/calj}

To appear in the Oct.\ 2015 issue of {\em Publications of the Astronomical Society of the Pacific} as a tutorial article.\\
Submitted: 2015-05-04. Revised: 2015-07-04. Accepted: 2015-07-08.

\section*{Abstract}

Astrometric surveys such as Gaia and LSST will measure parallaxes for hundreds of millions of stars. Yet they will not measure a single distance.  Rather, a distance must be estimated from a parallax.  In this didactic article, I show that doing this is not trivial once the fractional parallax error is larger than about 20\%, which will be the case for about 80\% of stars in the Gaia catalogue. Estimating distances is an inference problem in which the use of prior assumptions is unavoidable.  I investigate the properties and performance of various priors and examine their implications.  A supposed uninformative uniform prior in distance is shown to give very poor distance estimates (large bias and variance). Any prior with a sharp cut-off at some distance has similar problems.  The choice of prior depends on the information one has available -- and is willing to use -- concerning, for example, the survey and the Galaxy. I demonstrate that a simple prior which decreases asymptotically to zero at infinite distance has good performance, accommodates non-positive parallaxes, and does not require a bias correction.

\section{Introduction}

The parallax of an object is its observed angular displacement with respect to a reference frame due to the movement of the observer over a baseline. Stellar parallaxes are measured using twice the Earth--Sun separation as a baseline. From geometry and the definition of the parallax, $\parallax$, the distance of a star from the Sun, $r$\,pc, is $1/\parallax$\,\arcsec\ to a very good approximation. 

Parallaxes are one of the few distance measures in astronomy which do not make assumptions about the intrinsic properties of the object. Hipparcos blazed the trail when it measured the parallaxes of over $10^5$ stars to an accuracy of around 1\,mas down to $V \simeq 11$ in the 1990s (Perryman et al.\ 1997). Gaia is currently extending this to $10^9$ objects with expected accuracies as good as 0.01\,mas (and still much less than a mas at its magnitude limit of $G=20$) (Lindegren et al.\ 2008), and LSST hopes to obtain parallaxes with standard deviations of order 1\,mas for billions of stars down to much fainter magnitudes (Ivezi{\'c} et al.\ 2011).

It is therefore of considerable importance to know how to estimate distances (and their uncertainties) from parallaxes. Despite the simple relation between the two, inverting a parallax to give a distance is only appropriate when we have no measurement errors. As we always have measurement errors, determining the distance given a parallax becomes an inference problem.  Here I show that this inference is not trivial when the errors are even still quite small, necessarily involves assumptions, and can lead to surprisingly large errors.  I then set out to solve the following problem: given a measured parallax, $\parallax$, and its uncertainty, $\parsd$, how can we obtain a good estimate of the distance and its uncertainty?

Numerous studies in the astronomical literature have examined estimating physical quantities from parallaxes and possibly additional data (such as colours). These quantities may be individual distances, absolute magnitudes, distances to clusters, etc. Example studies include Lutz \& Kelker 1973, Smith \& Eichhorn 1996, Brown et al.\ 1997, Verbiest et al.\ 2012, Palmer et al.\ 2014. 
My objective is not to repeat such analyses, but rather to illustrate the issues involved using what is arguably the simplest of these problems: estimating distance from a single parallax. More complex problems will need to consider these same issues, even though in many cases we will not want to estimate distances explicitly. We should instead infer the quantity of interest directly. If we want to compare model predictions with data (which is not the topic of this paper), then it is usually better to project the predictions (and their distributions) into the data space, where the noise (measurement) model is better defined, rather than inferring quantities and then trying to determine an appropriate noise model to use in the comparison.

\subsection{Definitions}

I use $\varpi$ to denote parallax (always in \arcsec) and $r$ to denote distance (always in pc). Where it is necessary to make a distinction, I use $\rtrue$ to indicate the true distance, and $\rest$ to indicate an estimate of this, such as the mode $\rmode$, or median, $\rmed$.  I will frequently refer to the {\em fractional parallax error}, the ratio of the (1$\sigma$) parallax uncertainty to the parallax. This can either be the empirical quantity based only on the measured data, $\fpe = \parsd/\parallax$, 
or a quantity based on the true distance, $\fpetrue = \parsd \rtrue$.  In a real application $\rtrue$ and $\fpetrue$ are of course unknown.

\section{First steps}

\subsection{Measurement model (likelihood)}\label{sec:likelihood}

For a star at true distance $r$, its true but unknown parallax is $1/r$. The measured parallax, $\parallax$, is a noisy measurement of $1/r$. I will assume that $\parallax$ is normally distributed with unknown mean $1/r$ and known standard deviation $\parsd$. That is, I assume $\parallax$ has been drawn from the distribution
\begin{equation}
\like \,=\, \frac{1}{\sqrt{2 \pi}\parsd} \exp{ \left[ -\frac{1}{2\parsd^2}\left(\parallax-\frac{1}{r}\right)^2 \right] } \hspace*{1em}{\rm where}\hspace*{1em} \parsd \geq 0
\label{eqn:likelihood}
\end{equation}
which is Gaussian in $\parallax$, but of course not in $r$. 
This function is the measurement model, or likelihood. It gives the probability density function (PDF) -- probability per unit parallax -- for any $\parallax$, given values of $r$ and $\parsd$. 

This is the measurement model used in the Hipparcos and Gaia data processing. $\parsd$ depends in particular on the centroiding accuracy of each of the individual position measurements which are used in the astrometric solution. This in turn depends primarily on the number of photons ($N$) received from the star ($\parsd \simeq 1/\sqrt{N}$), 
and thus on its brightness, the observing time per observation, and the number of observations (assuming the observations have an appropriate cadence to determine a parallax at all). Depending on the source brightness, other terms and systematic errors may also be significant.
The two most important points here are (1) $\parsd$ is independent of $\parallax$, and (2) $\parsd$ can be estimated from another measurement model using other measured data (de Bruijne 2012).\footnote{As this measurement model is an approximation and uses noisy measurements of the star's brightness, it will not return the true value of $\parsd$. But this will generally be a small uncertainty compared to the noise in the parallax, so I don't consider it here.}

Equation~\ref{eqn:likelihood} has a finite probability of giving non-positive parallaxes, and this probability gets larger with increasing $\fpetrue$. An astrometric reduction can indeed produce a negative parallax, because the ``measured'' parallax is the result of reducing a set of noisy angular measurements. It corresponds to the parallactic displacement being in the opposite direction on the sky from that expected due to the movement of the observer along the baseline. It does {\em not} correspond to a negative distance, because $r>0$ by definition. Non-positive parallaxes are perfectly reasonable measurements, and we will see that they can deliver useful distance information.

Strictly speaking the likelihood must be changed at extremely small distances, because then the definition that the (noise-free) parallax is the reciprocal of distance breaks down. But this only occurs for parallaxes on degree scales, whereas all measured stellar parallaxes are less than an arcsecond.

\subsection{Intuition}\label{sec:intuition}

Suppose we have a measurement $\parallax \pm \parsd$. 

One may be tempted to report $1/\parallax$ as the distance estimate, and use a first order Taylor expansion to give $\parsd/\parallax^2$ as the uncertainty. But we will now see that this quickly becomes a poor approximation.

From the definition of the Gaussian, the intervals 
\begin{equation}
1/r \,=\, [\parallax-2\parsd, \parallax] \hspace{1em}{\rm and}\hspace{1em} 1/r \,=\, [\parallax, \parallax+2\parsd]
\end{equation}
each includes 
a fraction $0.954/2 = 0.477$ of the total probability of the distribution $\like$. The transformation from $1/r$ to $r$ is monotonic and so preserves the probability. Thus the intervals
\begin{equation}
r \,=\, [1/\parallax, 1/(\parallax-2\parsd)] \hspace{1em}{\rm and}\hspace{1em} r \,=\, [1/(\parallax+2\parsd), 1/\parallax]
\end{equation}
must each also contain a fraction 0.477 of the total probability over the distance. But whereas these intervals are equally sized in $1/r$ (the Gaussian is symmetric), they are not in $r$. For example, with $\parallax=0.1$ and $\parsd=0.02$, these intervals are $r=[10, 16.67]$ and $r=[7.14, 10]$ respectively. The uncertainties do not transform symmetrically because of the nonlinear transformation from $1/r$ to $r$. We see that the first order Taylor expansion suggested above is a poor approximation even for relatively small errors (here $\fpe$\,=\,$1/5$).

But what if the errors are larger, with $\fpe = 1/2$? The upper distance interval in the example becomes $r = [1/\parallax, \infty]$. If $\fpe$ is even larger, then this interval becomes undefined and we seem to ``lose'' some of the probability. As the Gaussian distribution has infinite support for all values of $\parallax$ and $\parsd$, some finite amount of probability in the likelihood function will always correspond to an undefined distance.

The problem here is that we are trying to make a probabilistic statment about $r$ using just equation~\ref{eqn:likelihood}, yet this is a probability distribution over $\parallax$, not $r$. The solution is to pose the problem correctly.

\section{The inference problem}

Given $\parallax$ and $\parsd$, we want to infer $r$. As there is noise involved, we know we cannot infer $r$ exactly, so we want a probability distribution over the possible values of $r$. That is, we want to find $\post$. From Bayes theorem (which follows from the axioms of probability) this is related to the likelihood by
\begin{equation}
\post \,=\,\frac{1}{Z}\,\like\,\prior
\label{eqn:bayes}
\end{equation}
where $Z$ is the normalization constant
\begin{equation}
Z \,=\, \int_{r=0}^{r=\infty} \like\,\prior\,dr
\end{equation}
which is not a function of $r$. 
$\prior$ is the prior, and expresses our knowledge of -- or assumptions about -- the distance, independent of the parallax we have measured. $\post$ is the posterior distribution. 
By multiplying the likelihood by a prior, we essentially transform an expression (the likelihood) for the probability of the known data (parallax) given the unknown parameter (distance) to an expression (the posterior) for the probability of the parameter given the data.

We see that we can only infer the distance in a probabilistic sense if we adopt a prior. Some people object to this on philosophical grounds ({\em How can science depend on assumptions?}), others on practical grounds ({\em How can I know the prior if I haven't yet measured any distances?}). The latter is a valid objection and will be discussed later. Yet without a prior, we run into the problems we just saw in the previous section.

\section{An improper uniform prior}\label{sec:runif_improper}

A common strategy for dealing with prior discomfort is to adopt a uniform prior over the full range of the parameter, on the grounds that this does not prefer one value over another. In the present context we should go one step further and use
\begin{equation}
  \uprior[iu]  \,=\,  \begin{dcases}
  1  & \:{\rm if}~~ r > 0 \\
  0                          & \:{\rm otherwise}
\end{dcases}
\label{eqn:urunifprior}
\end{equation}
so as to introduce the definition that distances must be positive.
The $^*$ symbol is used to indicate that the PDF is unnormalized. The subscript is an abbreviation of the distribution. Because it extends to infinity, this prior cannot be normalized. Such priors are referred to as {\em improper}.
From equation \ref{eqn:bayes}, the
(unnormalized) posterior, $\upost[iu]$, in this case is the likelihood (equation~\ref{eqn:likelihood}) but now considered as a function of $r$ rather than $\parallax$, and subject to the additional constraint
that $r$ must be positive, i.e.\
\begin{equation}
\upost[iu]  \,=\,  \begin{dcases}
  \like  & \:{\rm if}~~ r > 0 \\
  0                          & \:{\rm otherwise.}
\end{dcases}
\label{eqn:urunifpost}
\end{equation}

\begin{figure}
\begin{center}
\includegraphics[width=0.50\textwidth, angle=0]{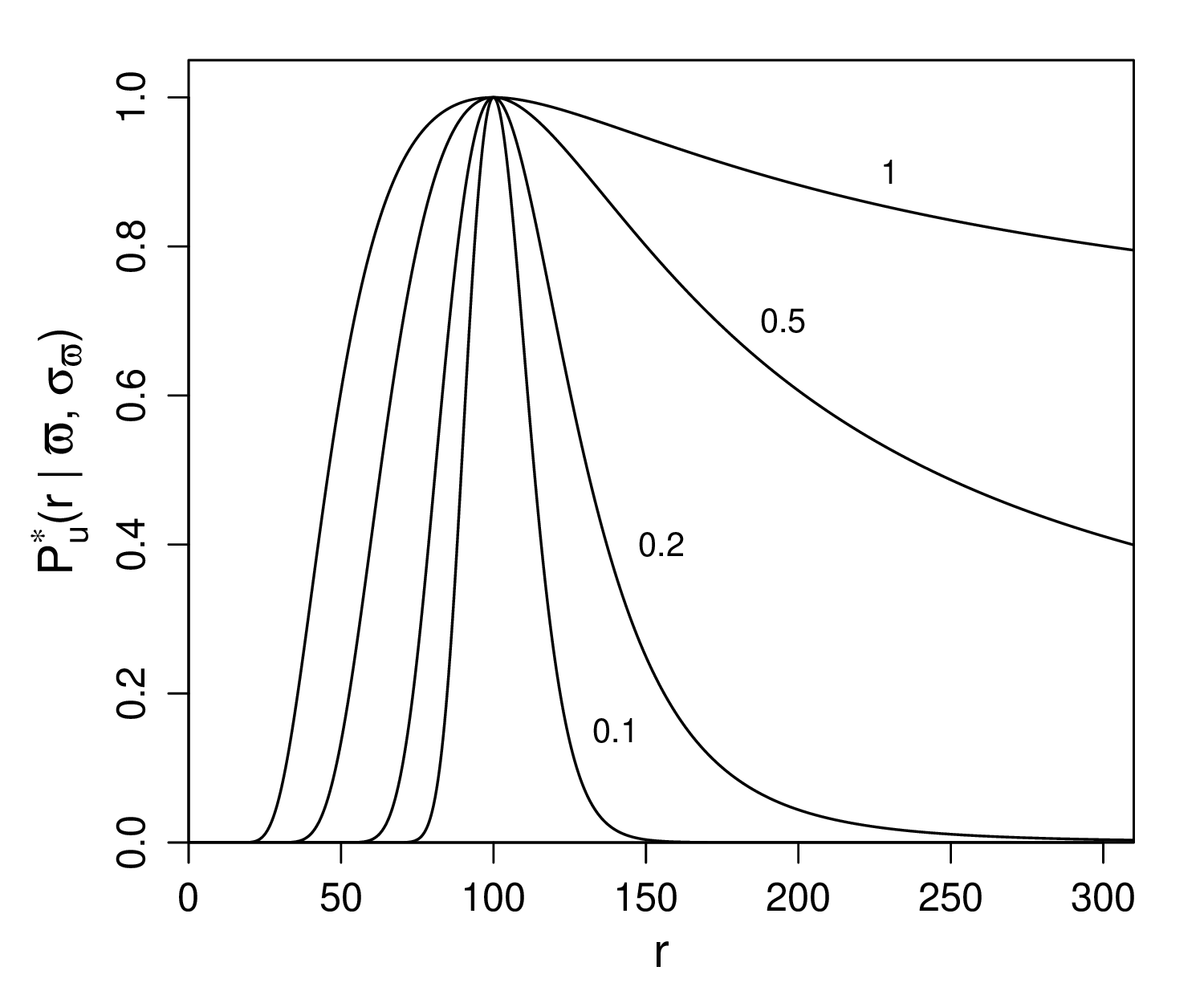}
\caption{The unnormalized posterior $\upost[iu]$ (improper uniform
  prior) for $\parallax=1/100$ and four values of $\fpe=(0.1, 0.2,
  0.5, 1.0)$. The posteriors have been scaled to all have their mode
  at $\upost[iu]=1$. As the prior is improper, the posterior remains finite out to infinite distance for all values of $\fpe>0$.\label{fig:ud.post_runifPrior}}
\end{center}
\end{figure}

Examples of this posterior are shown in Figure \ref{fig:ud.post_runifPrior} for $\parallax=1/100$ and various values of $\fpe$. This demonstrates the skewness discussed in section \ref{sec:intuition}.

Inspection of equation \ref{eqn:likelihood} shows that
\begin{equation}
\lim_{r \to \infty} \upost[iu] \,=\, {\rm const} \ .
\end{equation}
The posterior does not converge, has an infinite area and so cannot be normalized. Consequently it has no mean, no standard deviation, no median, and no quantiles. The only plausible estimator of the distance is the mode\footnote{As the prior is uniform here, the maximum of the posterior equals the maximum of the likelihood when the latter is expressed as a function of $r$.}, which we see from Figure~\ref{fig:ud.post_runifPrior} is well defined for all values of $\fpe$, and is equal to $1/\parallax$ for $\parallax > 0$.  
For non-positive parallaxes the posterior increases monotonically from 0 at $r=0$ to a finite asymptote
(equation \ref{eqn:likelihood}) putting the mode at $r = \infty$, which is physically implausible in a finite universe.
This is bad, as negative parallaxes are valid measurements.
We should also be concerned that this estimator ignores the parallax uncertainty, even though we have seen that its magnitude determines the skewness of the distance distribution (section \ref{sec:intuition}).  
An uncertainty could be derived from the full width at half maximum (FWHM) (or similar), but without a normalizable posterior there is no useful probabilistic interpretation of this.

\subsection{Empirical test}\label{sec:empirical_test}

The above mentioned problems notwithstanding, we can still ask how good the mode of this posterior is as an estimator. Recall that the posterior is a function of the {\em measured} parallax, which due to noise is not equal to the true parallax, so $1/\parallax$ is not equal to the true distance.
Thus to assess the quality of any estimator we need to test it using noisy simulated data. In particular, we would like to see how its accuracy varies as a function of the expected fractional parallax error.
I do this with the following empirical test procedure:
\vspace*{-0.5em}
\begin{myenumerate}
\item assign a fractional parallax error, $\fpetrue$
\item assign a true distance, $\rtrue$ (which together with $\fpetrue$ determines $\parsd$) \label{listitem:assign_distance}
\item simulate a parallax measurement, $\parallax$, by drawing a value at random from the likelihood with $r=\rtrue$\label{listitem:draw_parallax}
\item use $\parallax$ in the posterior to evaluate the distance estimator, $\rest$
\item calculate the {\em scaled residual}, $x = (\rest - \rtrue)/\rtrue$
\item repeat steps 2--5 numerous times to compile a set of values $\{x_i\}$ at a single value of $\fpetrue$
\item calculate the bias, $b(\fpetrue) = {\overline x}$, and sample standard deviation, $s(\fpetrue) =
\sqrt{\frac{1}{n-1} \sum_i (x_i - \overline{x})^2}$
\item repeat steps 1--7 for different values of $\fpetrue$.
\end{myenumerate}
\vspace*{-0.5em}
The bias and standard deviation are standard tests of the quality of an estimator. A good estimator will have small values of both, where ``small'' is, of course, a relative term.
By using the scaled residuals (rather than just the residuals) I can collate information on many different true distances into a single value of $b$ and $s$ for a given $\fpetrue$.

In the first test I assign true distances in step \ref{listitem:assign_distance} by drawing at random from the distribution
\begin{equation}
\prior[r^2]  \,=\,  \begin{dcases}
  \frac{3}{\rlim^3}\,r^2  & \:{\rm if}~~ 0 < r \leq \rlim \\
  0                          & \:{\rm otherwise}
\end{dcases}
\label{eqn:r2prior}
\end{equation}
which corresponds to stars being uniformly distributed in three-dimensional space,
$P(V)$\,=\,const.\footnote{For a constant volume density the probability, $P(V)dV$, of finding a star in a shell with inner and outer radii $(r, r + dr)$ is proportional to the volume of that shell, $4\pi r^2 dr$. $P(V)dV=P(r)dr$, so $P(r) \propto r^2$.} I use $\rlim=10^3$. For each value of $\fpetrue$ I sample $10^6$ true distances. I do this for 100 values of $\fpetrue$ ranging from 0.01 to 1.0 in steps of 0.01.
The estimator in this case is the mode, which is just $1/\parallax$.
If the parallax is non-positive (in step \ref{listitem:draw_parallax}), then for the improper uniform prior I throw away the sample, because an infinite
distance would give an infinite bias and standard deviation (telling us straight away that it is a bad estimator!).
The procedure for other priors I will discuss as we encounter them.

\begin{figure}
\begin{center}
\includegraphics[width=0.49\textwidth, angle=0]{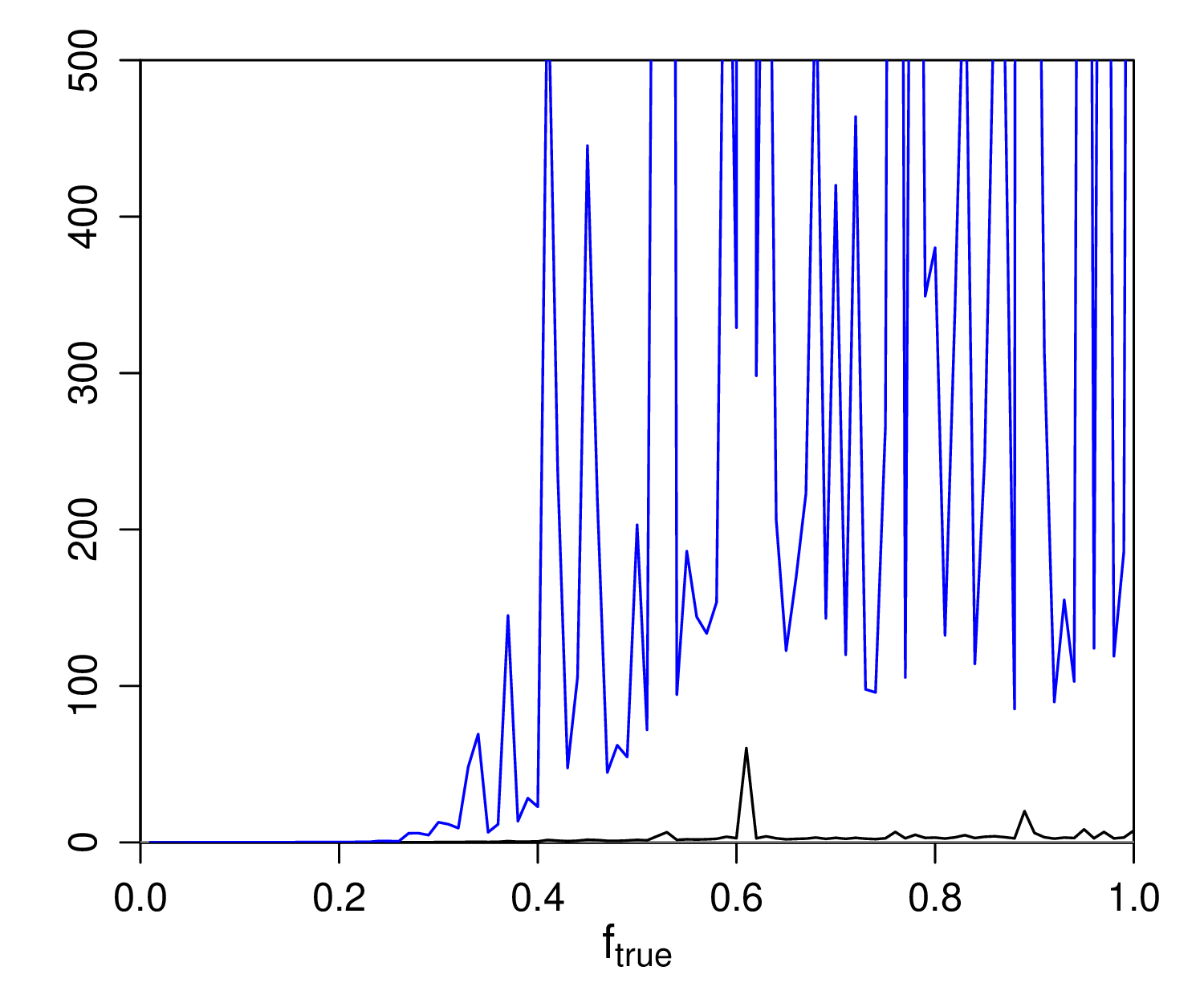}
\includegraphics[width=0.49\textwidth, angle=0]{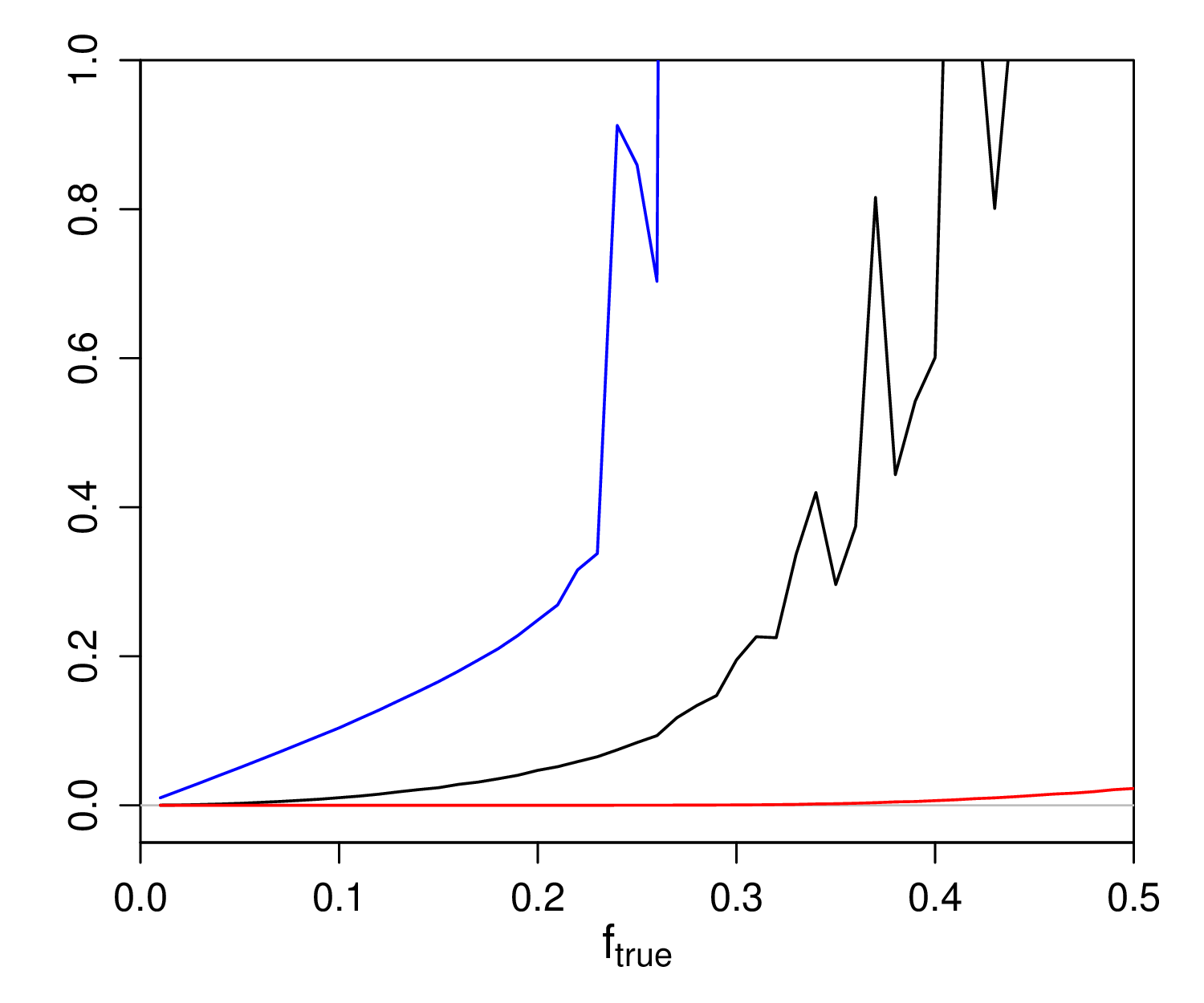}
\caption{The bias (black) and standard deviation (blue) as a function of $\fpetrue$ for
the mode estimator of the posterior $\upost[iu]$ (equation \ref{eqn:urunifpost}, which uses
the improper uniform prior) for data drawn from the
constant volume density prior (equation~\ref{eqn:r2prior}). The right panel is a zoom of the left panel. The red line in the right panel shows the fraction of samples which had to be rejected because they had non-positive parallaxes. \label{fig:scaledResiduals_mode_runifPrior_r2TruePrior_rmax1e3}}
\end{center}
\end{figure}

The results are shown in Figure \ref{fig:scaledResiduals_mode_runifPrior_r2TruePrior_rmax1e3}. We see that the standard deviation increases steadily until $\fpetrue \simeq 0.22$, then it explodes. The bias also increases sharply, in this case reaching a value of about 4 at $\fpetrue$\,=\,1. The reason for this appearance is that as $\fpetrue$ increases, the probability increases that the measured parallax -- which is drawn from the likelihood -- becomes arbitrarily close to zero, in which case $\rest=1/\parallax$ becomes arbitrarily large.  
This also explains why both the bias and standard deviation become very noisy for $\fpetrue \gtrsim 0.22$: the exact appearance of the curves depends on the sample drawn in the simulation.\footnote{The definition of the point at which the standard deviation, and especially the bias, ``explodes'', is somewhat arbitrary, but it appears to be reasonably stable for sufficiently large samples. Furthermore, 
repeating the experiment with larger samples, it seems that the point at which the standard deviation ``explodes'' converges to $\fpetrue \simeq 0.20$. Below this value we get a very smooth dependence of bias and standard deviation on $\fpetrue$. At values of
$\fpetrue$ well below this ``explosion'' we could even use the Taylor approximation mentioned in section \ref{sec:intuition} to estimate the distance and a (symmetric) uncertainty.}

Although noisy, the plot is independent of $\rlim$, because we are calculating functions of the scaled residuals, $x$, for fixed $\fpetrue$. Doubling $\rlim$ would double both $\rtrue$ and $\rest$, leaving $x$ unchanged.

\begin{figure}
\begin{center}
\includegraphics[width=0.49\textwidth, angle=0]{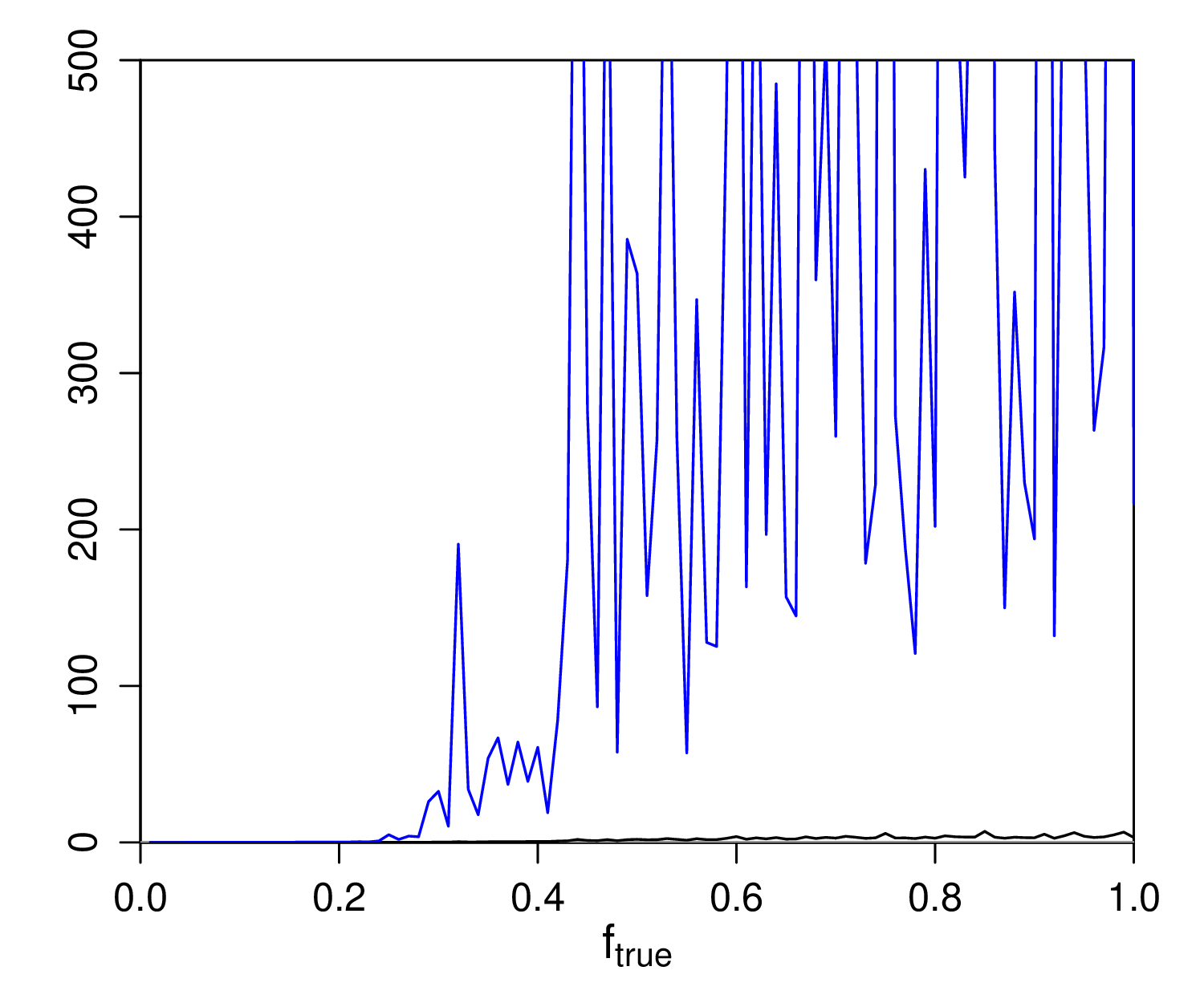}
\includegraphics[width=0.49\textwidth, angle=0]{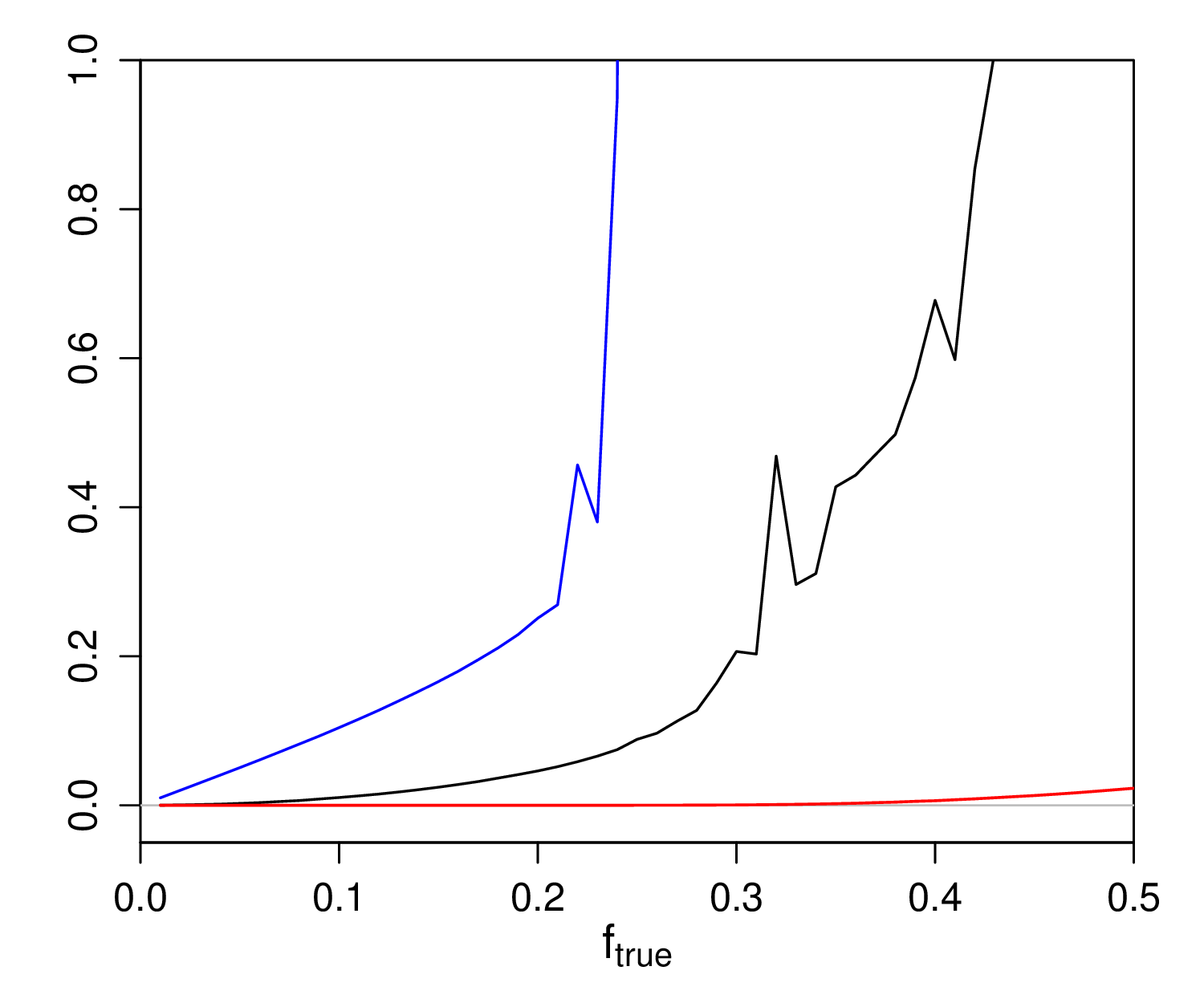}
\caption{As Figure~\ref{fig:scaledResiduals_mode_runifPrior_r2TruePrior_rmax1e3}, but now drawing data from the truncated uniform prior.\label{fig:scaledResiduals_mode_runifPrior_rmax1e3}}
\end{center}
\end{figure}

One might argue that by drawing true distances from a distribution which is very different from the uniform prior adopted in the posterior, we are bound to get poor results. To some extent this is true. To test this, I repeat the above procedure, but drawing from the prior in equation \ref{eqn:urunifprior}. Of course, to be able to draw from this I have to set an upper limit (otherwise all draws will be infinite), and I again use $\rlim=10^3$ (although the exact value is irrelevant, because the scaled residuals are independent of it).
The results are shown in Figure \ref{fig:scaledResiduals_mode_runifPrior_rmax1e3}. We see that they are hardly any better than before. The problem is therefore not prior mismatch, but the fact that $1/\parallax$ (the mode) is a very noisy estimator once $\fpetrue \gtrsim 0.22$. Because this posterior is not normalizable, no other estimator is available. 

This example shows that an apparently innocuous approach to inference is in fact highly problematic.  It is the kind of thing which may be done by those who reject the Bayesian approach to inference on the grounds that it depends on prior assumptions. They want to rely only on the likelihood, yet have to adopt some kind of prior to get logically consistent answers, and tacitly choose an unbounded uniform prior in $r$ in the belief that it is ``uninformative''.  Yet one can equally argue that uniform in $\log r$ (i.e.\ in distance modulus) is uninformative.  In fact, by adopting a uniform prior in $r$, one assumes that the volume density of stars varies as $P(V) \sim 1/r^2$ (see the discussion around equation \ref{eqn:r2prior}), which is highly informative.\footnote{Uniform in $\log r$ corresponds to uniform in $1/r$ and so $P(V) \sim 1/r^3$.}

Clearly, this approach is not very robust: it will only give distance estimates for $\fpetrue \lesssim 0.22$, it cannot give self-consistent uncertainty estimates (only a poor approximation using a Taylor expansion), and it breaks down entirely with non-positive parallaxes. Can we do better?

\section{A proper uniform prior}\label{sec:runif_proper}

\begin{figure}
\begin{center}
\includegraphics[width=0.49\textwidth, angle=0]{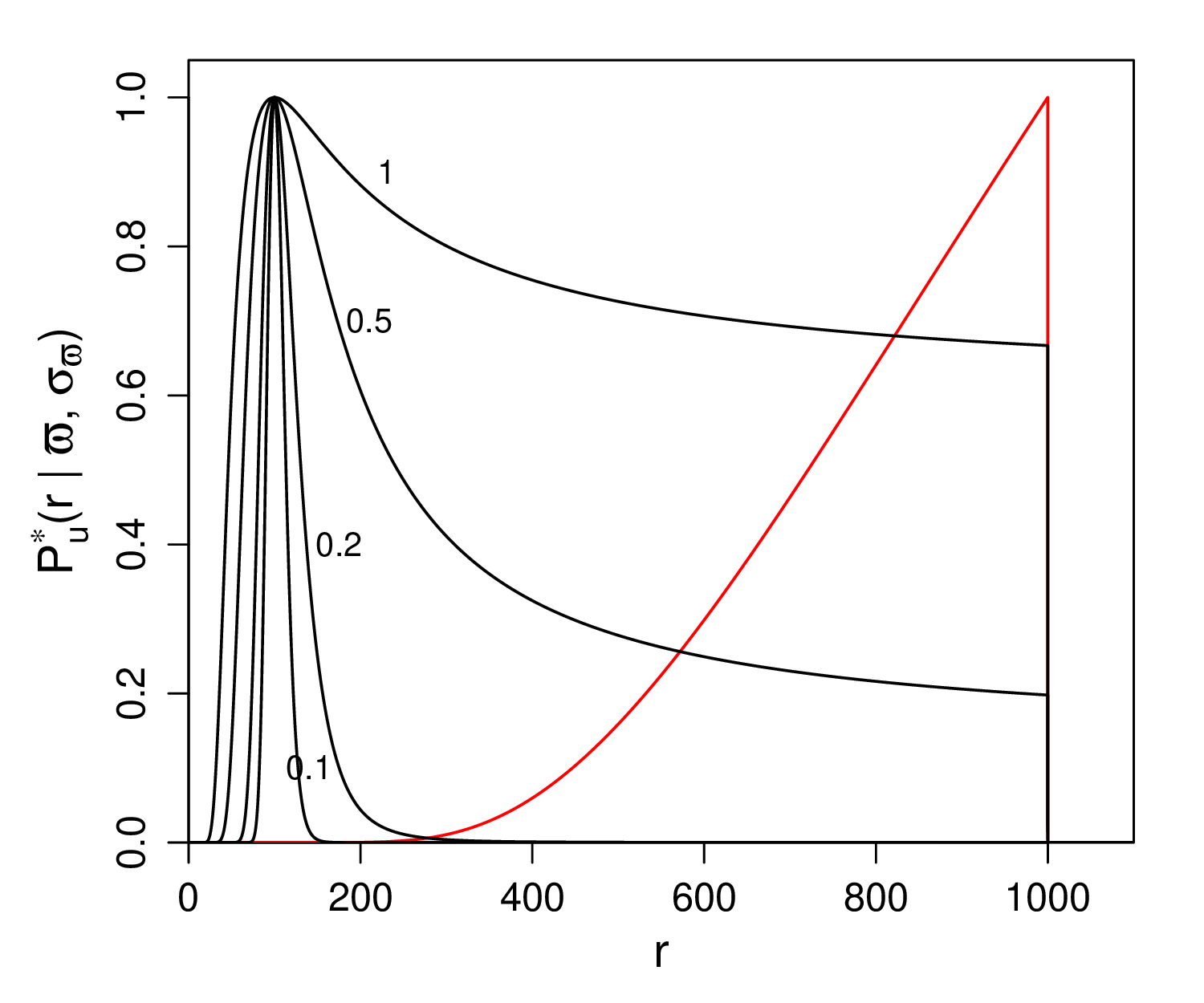}
\includegraphics[width=0.49\textwidth, angle=0]{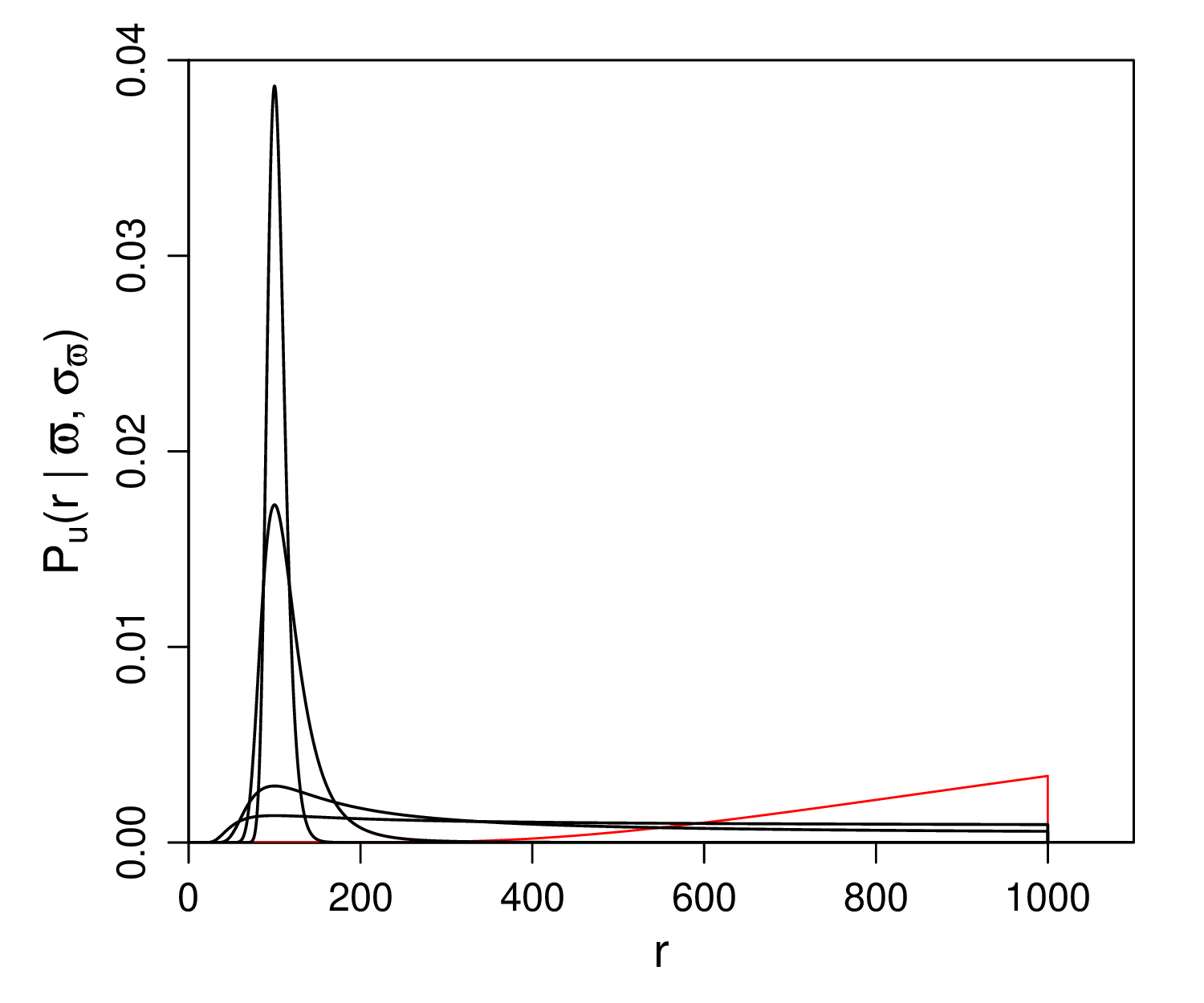}
\caption{Left: the unnormalized posterior $\upost[u]$ (truncated uniform prior with $\rlim=10^3$) 
for $\parallax=1/100$ and four values of $\fpe=(0.1, 0.2, 0.5, 1.0)$ (black lines). 
The red line shows the posterior for $\parallax = -1/100$ and $|\fpe|=0.25$.
The posteriors have been scaled to all have their mode at 
at $\upost[u]=1$. Right: the same five posterior PFDs but now normalized.
\label{fig:ud.post_runiftruncPrior}}
\end{center}
\end{figure}

An obvious improvement is to truncate the prior at some value. 
This prevents the inferred distance becoming very large, so should reduce the explosion of the bias and standard deviation we saw with the improper prior for larger $\fpetrue$. The normalized prior is
\begin{equation}
\prior[u]  \,=\,  \begin{dcases}
  \frac{1}{\rlim}  & \:{\rm if}~~ 0 < r \leq \rlim \\
  0                          & \:{\rm otherwise}  \ .
\end{dcases}
\label{eqn:runifprior}
\end{equation}
where $\rlim$ is the largest distance we expect for any star in our survey.
The posterior is the same shape as in Figure \ref{fig:ud.post_runifPrior} but set to zero for $r>\rlim$
\begin{equation}
\upost[u]  \,=\,  \begin{dcases}
  \frac{1}{\rlim}\like  & \:{\rm if}~~ 0 < r \leq \rlim \\
  0                          & \:{\rm otherwise} \ .
\end{dcases}
\label{eqn:runifpost}
\end{equation}
This is shown in Figure \ref{fig:ud.post_runiftruncPrior}. Note the effect of the normalization.\footnote{This and all subsequent posteriors have no simple solutions for their integrals, so the integrals
have been evaluated numerically for the plots by sampling on a fine grid.}
The right panel illustrates how the posterior is a combination of the likelihood and prior.
When the data are good ($\fpetrue$ is small), the likelihood is narrow compared to the prior,
so the former dominates the posterior. As the data become less informative (larger $\fpetrue$),
the posterior gets broader and the prior plays more of a role.
For non-positive parallaxes the posterior increases 
monotonically from 0 at $r=0$ to a maximum at $r=\rlim$ (the red line shows an example).

The mode of the resulting posterior serves as our distance estimator, and is
\begin{equation}
  \rest  \,=\,  \begin{dcases}
  \frac{1}{\parallax}  & \:{\rm if}~~ 0 < \frac{1}{\parallax} \leq \rlim \\
  \rlim                   & \:{\rm if}~~ \frac{1}{\parallax} > \rlim  \hspace*{1em}{\rm (extreme~mode)} \\
  \rlim                   & \:{\rm if}~~ \parallax \leq 0 \ . \\
\end{dcases}
\label{eqn:postut_mode}
\end{equation}
I call a mode solution ``extreme'' if the estimate from the likelihood alone would place it beyond $\rlim$, and so is truncated by the prior to $\rlim$.
That non-postitive parallaxes are set to 
$\rlim$ is consistent with the nature of the parallax measurement (see section \ref{sec:likelihood}). 

\begin{figure}
\begin{center}
\includegraphics[width=0.49\textwidth, angle=0]{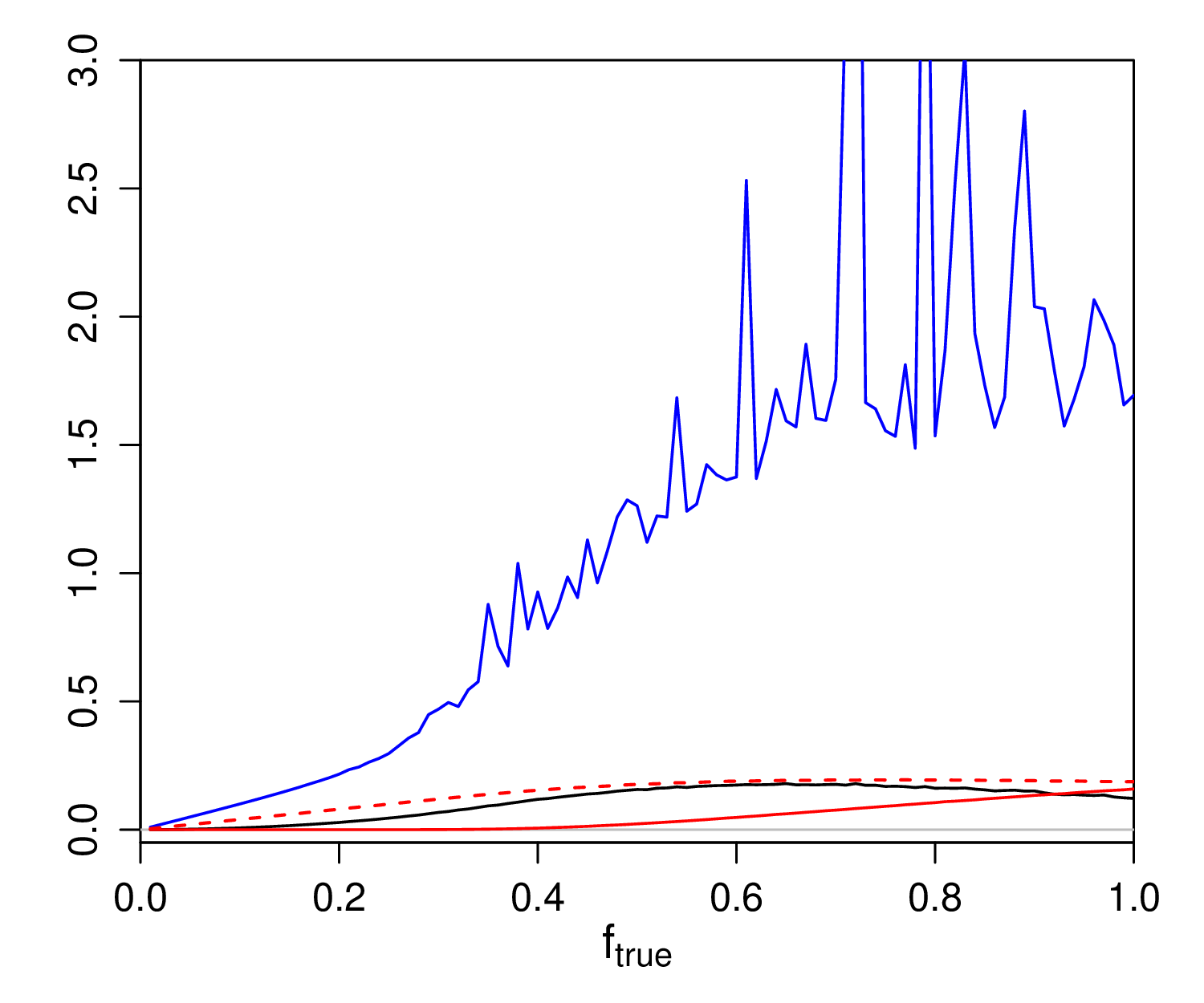}
\includegraphics[width=0.49\textwidth, angle=0]{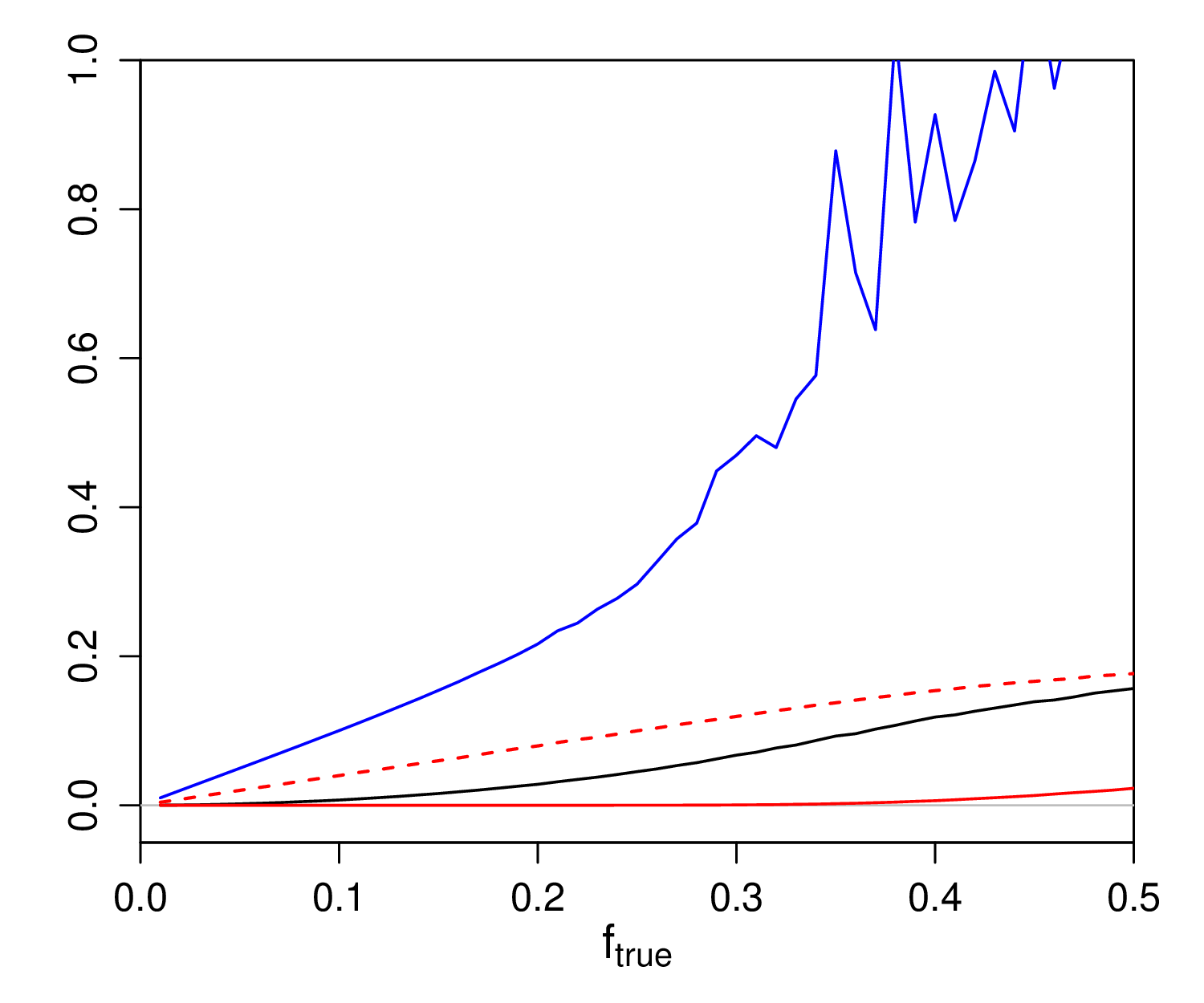}
\caption{The bias (black) and standard deviation (blue) as a function of $\fpetrue$ for
the mode estimator (equation \ref{eqn:postut_mode}) of the posterior $\post[u]$ (equation \ref{eqn:runifpost}, which uses the truncated uniform prior) for data drawn from a truncated uniform prior in $r$. The right panel is a zoom of the left panel. The solid red line shows the fraction of samples which were rejected because they had non-positive parallaxes. 
The dashed red line shows the fraction of samples with an extreme mode ($1/\parallax > \rlim$) which was truncated to $\rlim$. \label{fig:scaledResiduals_mode_runifPrior_rmax1e3_truncateMode}}
\end{center}
\end{figure}

\begin{figure}
\begin{center}
\includegraphics[width=0.49\textwidth, angle=0]{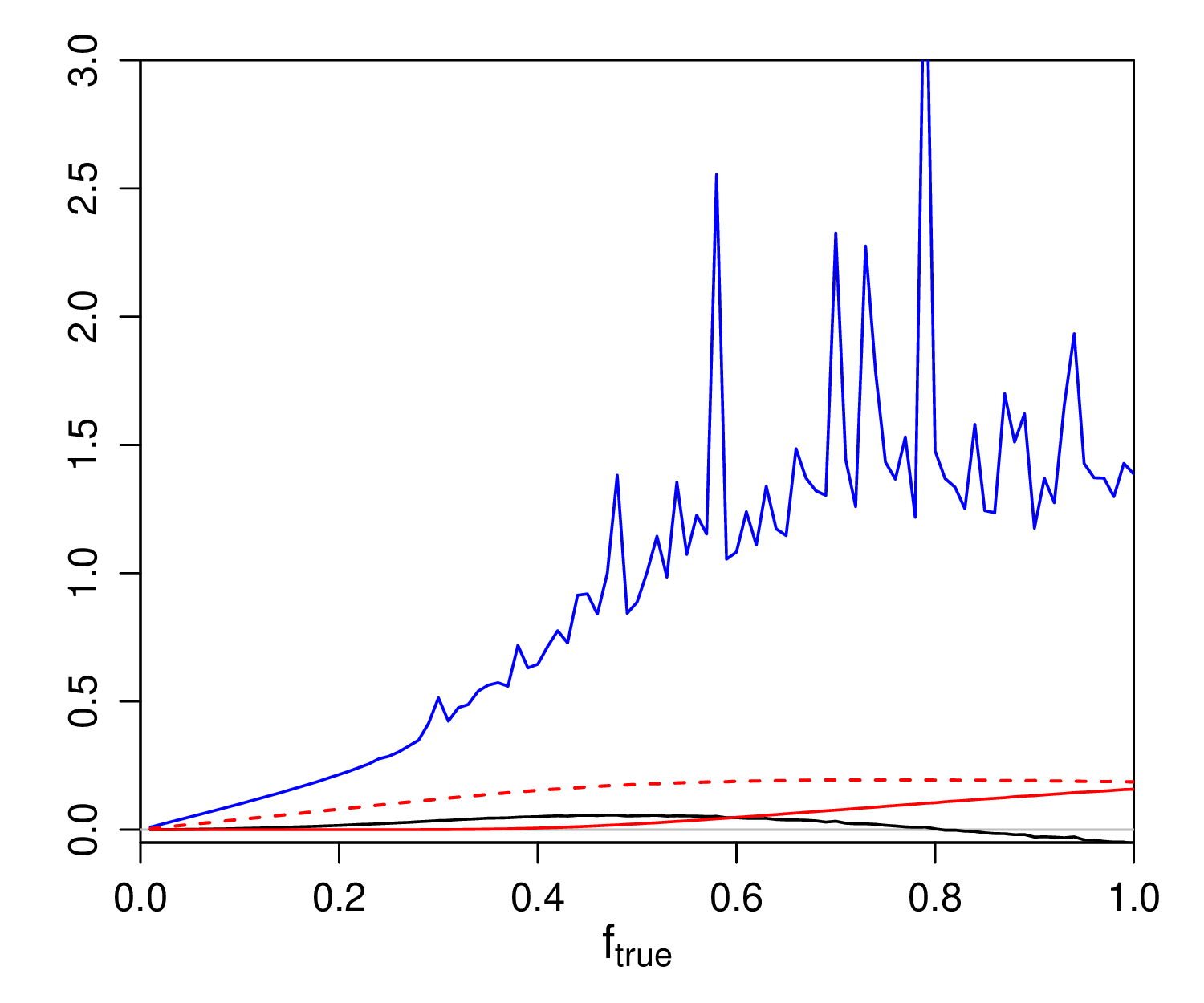}
\includegraphics[width=0.49\textwidth, angle=0]{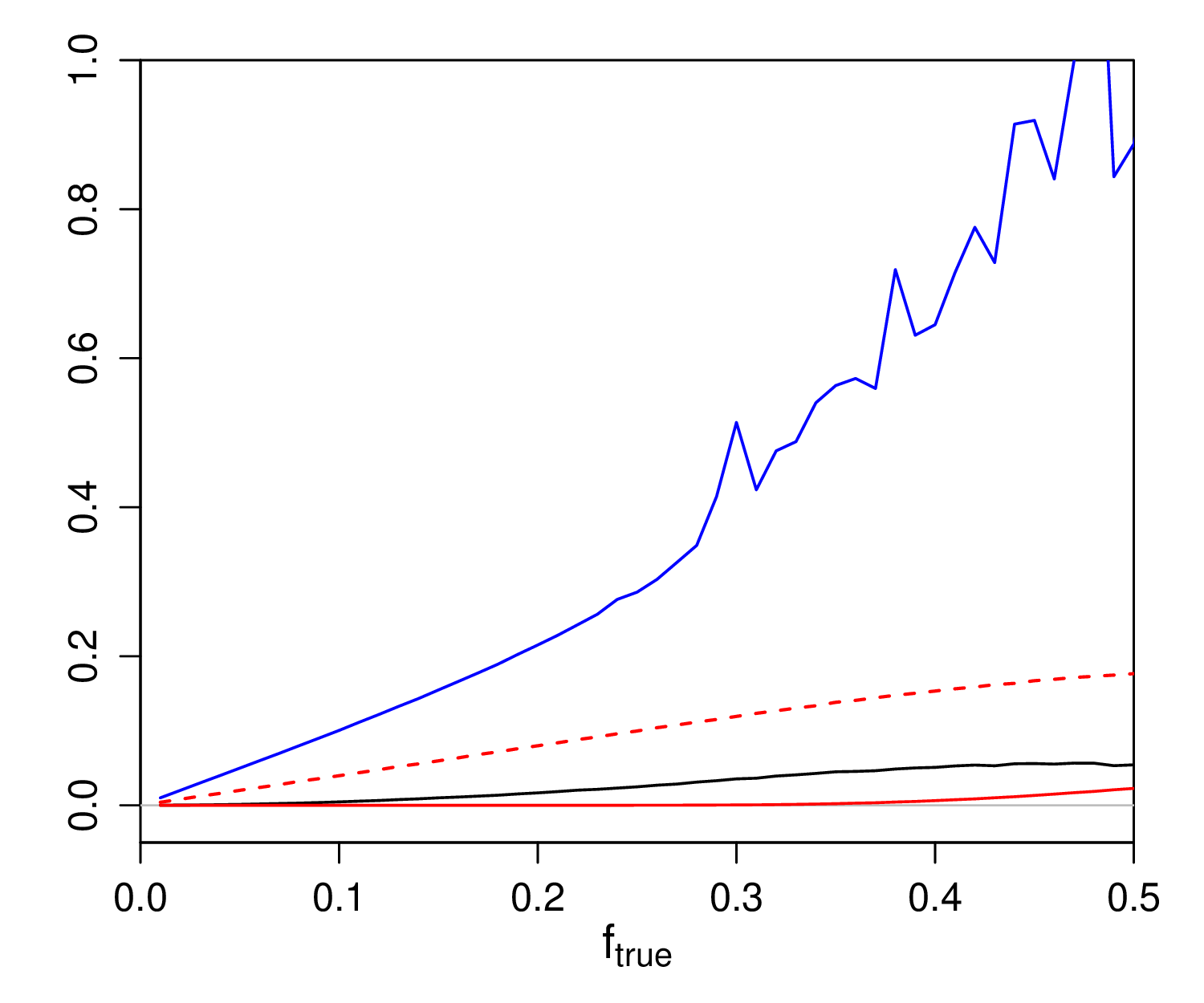}
\caption{As Figure \ref{fig:scaledResiduals_mode_runifPrior_rmax1e3_truncateMode}, but now rejecting all samples which have $1/\parallax > \rlim$. \label{fig:scaledResiduals_mode_runifPrior_rmax1e3_rejectMode}}
\end{center}
\end{figure}

I redo the empirical test with data drawn from the uniform prior (with a common value of $\rlim$).
To allow a fairer comparison with the results obtained with the improper uniform prior,
I again reject simulated parallax measurements in step \ref{listitem:draw_parallax} which are non-positive.
The results are shown in Figure \ref{fig:scaledResiduals_mode_runifPrior_rmax1e3_truncateMode}.  They are again independent of $\rlim$.  The bias and standard deviation are significantly reduced. This is not surprising, because we are now clipping all those large distance estimates to $\rlim$. The dashed line shows that relatively few objects have their distances clipped in this way -- only 10\% at $\fpetrue$\,=\,0.25 and 20\% at $\fpetrue$\,=\,1 -- which in turn shows that the explosion in the bias and standard deviation seen previously was due to a minority of objects. An alternative to clipping these distances would be to reject them, which means we decide we cannot estimate distances. The results of the empirical test in that case (for the same data) are shown in Figure \ref{fig:scaledResiduals_mode_runifPrior_rmax1e3_rejectMode}. As we would expect, rejecting these cases reduces the bias and standard deviation, but not by much.

Compared to the improper uniform prior at least, the results look quite good. In Figure \ref{fig:scaledResiduals_mode_runifPrior_rmax1e3_rejectMode}, the bias and standard deviation increase approximately linearly for $\fpetrue<0.25$ with gradients of 0.10 and 1.1 respectively; their values are 0.025 and 0.29 respectively at $\fpetrue=0.25$.  One might think about correcting the bias (see section \ref{sec:biascorrection}), but the need for this will be later obviated by the use of a better prior.

Now that the posterior is normalized we could investigate using the mean or median.  But these are strongly influenced by the choice of $\rlim$ for large values of $\fpetrue$, which is just the domain where we want a more robust estimator than the mode.

The results in Figures \ref{fig:scaledResiduals_mode_runifPrior_rmax1e3_truncateMode} and \ref{fig:scaledResiduals_mode_runifPrior_rmax1e3_rejectMode} are for the situation when the real data come from the same distribution as the prior.  The shape of this prior aside, individual distance estimates for small parallaxes and/or large values of $\fpetrue$ will be sensitive to the choice of $\rlim$.  Astrophysically this corresponds to knowing the maximum distance of any star. This could be set for each star individually according to its measured apparent magnitude, $m$, and the fact that stellar evolution limits the absolute magnitude, $M_{\rm lim}$, to about $-5$\,mag. The flux continuity relation then tells us that $5 \log_{\rm 10}\rlim = m - M_{\rm lim} - A_m + 5$, where $A_m$ is the interstellar absorption in the observed band.  Setting $A_m=0$ gives us the maximum distance (although this will be an overestimate for observations at low Galactic latitudes). A more stringent upper limit on $M_{\rm lim}$ could be set if the star's colour were known.

Truncating the prior has helped, but it still corresponds to assuming that the volume density of stars varies as $P(V) \sim 1/r^2$. This is not only a strong prior, it is also scale independent, which the Galactic stellar distribution most certainly is not.

\section{A constant volume density prior}

\begin{figure}
\begin{center}
\includegraphics[width=0.49\textwidth, angle=0]{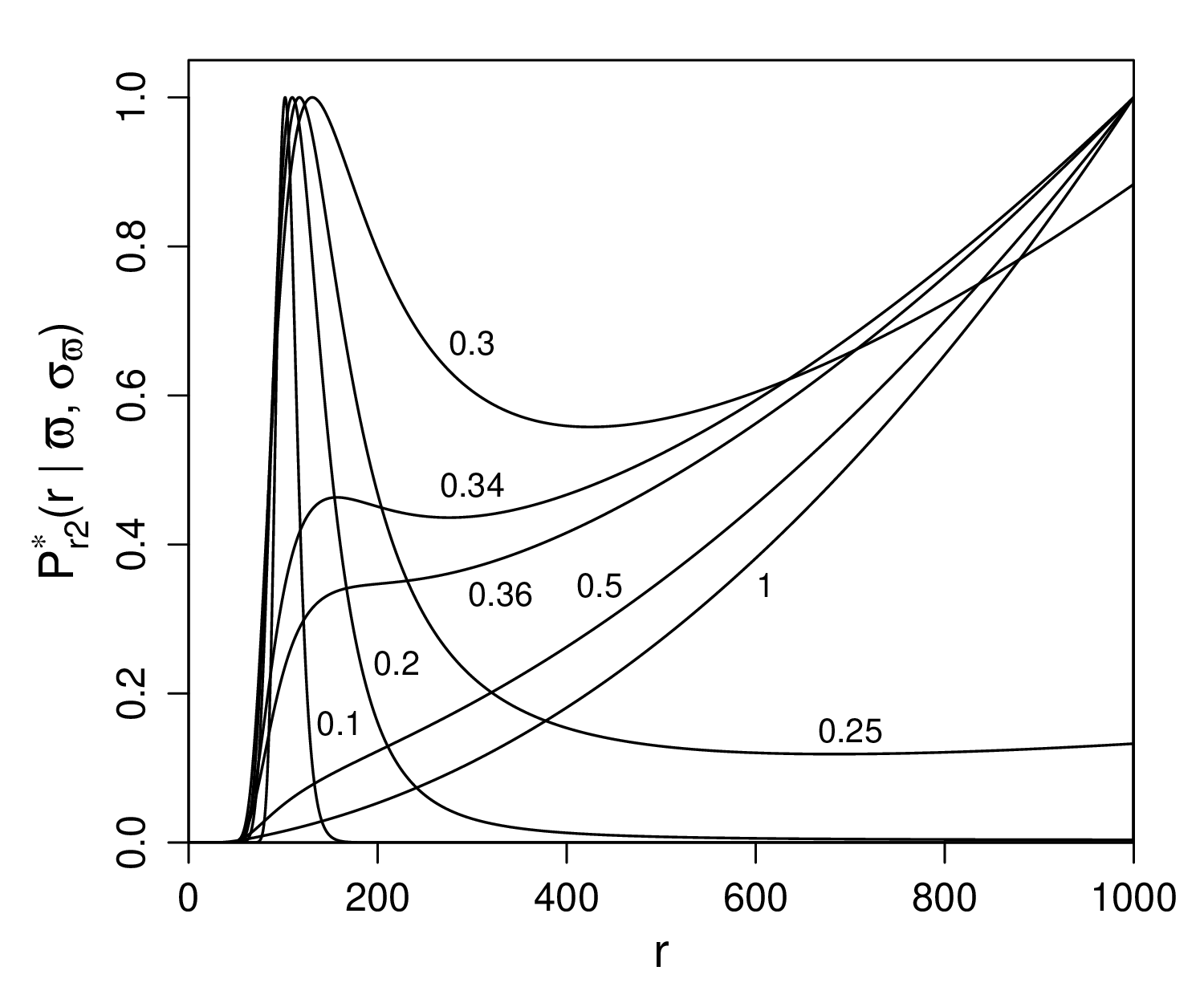}
\includegraphics[width=0.49\textwidth, angle=0]{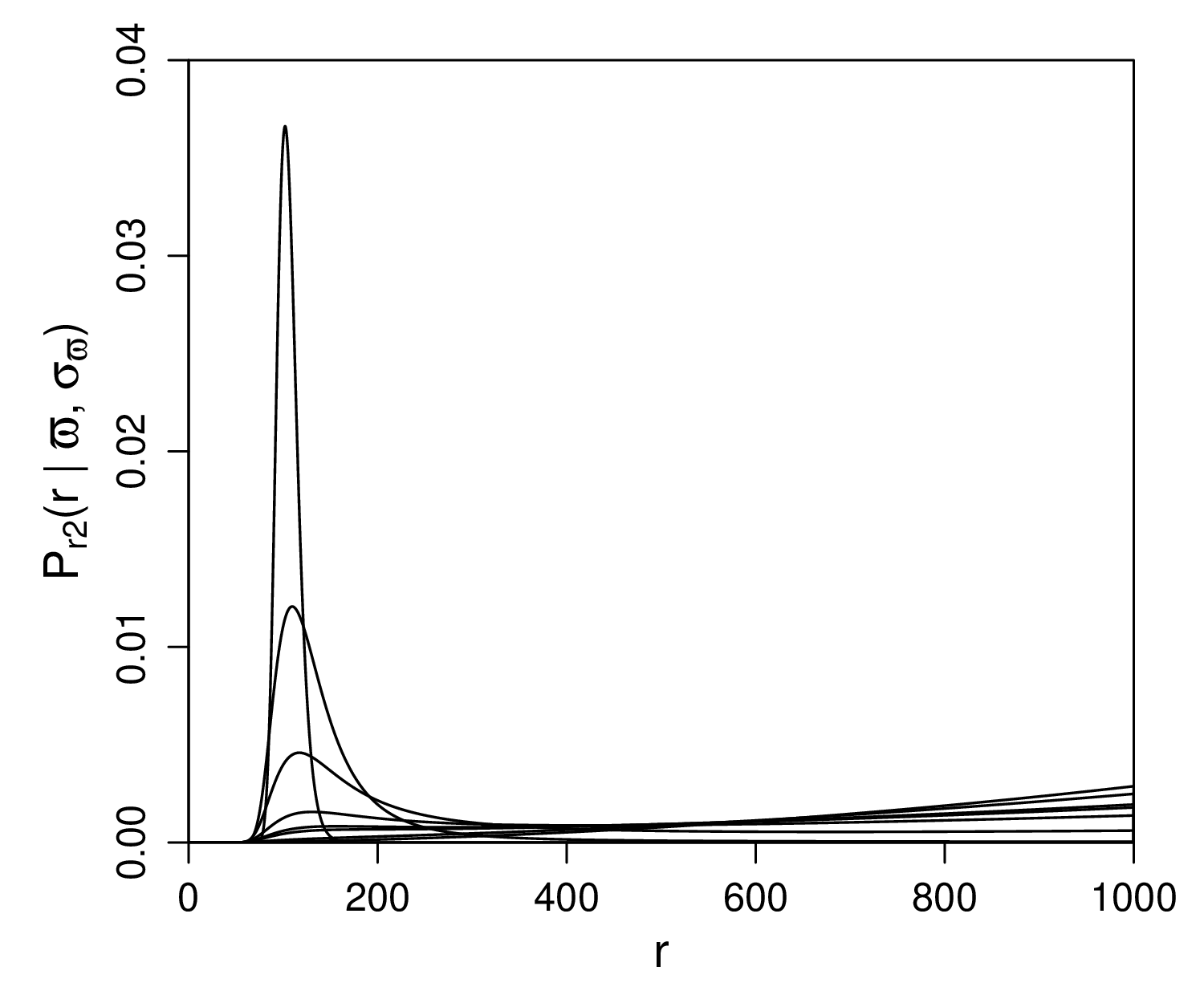}
\caption{Left: the unnormalized posterior $\upost[r^2]$ (truncated constant volume density prior with $\rlim=10^3$) for $\parallax=1/100$
and eight values of $\fpe=(0.1, 0.2, 0.25, 0.3, 0.34, 0.36, 0.5, 1.0)$.
The posteriors have been scaled to all have their mode at 
at $\upost[r^2]=1$. Right: the same posterior PDFs but now normalized. The curves with the clear maxima around $r=100$ are $\fpe=(0.1, 0.2, 0.25, 0.3)$ in decreasing order of the height of the maximum.\label{fig:post_r2Prior}}
\end{center}
\end{figure}

One could adopt a more complex prior based on a Galaxy model. We may want to do this in practice (see section \ref{sec:choice}), but this gives us little insight into the role of the prior. Let us instead take at least an astrophysically reasonable yet less informative prior, namely a constant volume density of stars (equation \ref{eqn:r2prior}). This too is truncated at $r=\rlim$ in order to be normalizable. The unnormalized posterior is
\begin{equation}
  \upost[r^2]  \,=\,  \begin{dcases}
  \frac{r^2}{\parsd} \exp{ \left[ -\frac{1}{2\parsd^2}\left(\parallax-\frac{1}{r}\right)^2 \right] }  & \:{\rm if}~~ 0 < r \leq \rlim \\
  0                          & \:{\rm otherwise} \ .
\end{dcases}
\label{eqn:upostr2}
\end{equation}
Examples of this posterior are shown in Figure \ref{fig:post_r2Prior} for $\parallax=1/100$, $\rlim=10^3$, and various values of $\fpe$. 
For low $\fpe$ the posterior looks unimodal, but then appears to develop a minimum (and thus a second mode at $r=\rlim$), until for $\fpe\gtrsim0.35$ the lower mode (and hence the minimum) disappears, leaving a posterior which increases monotonically from zero up to $\rlim$. 

\begin{figure}
\begin{center}
\includegraphics[width=0.98\textwidth, angle=0]{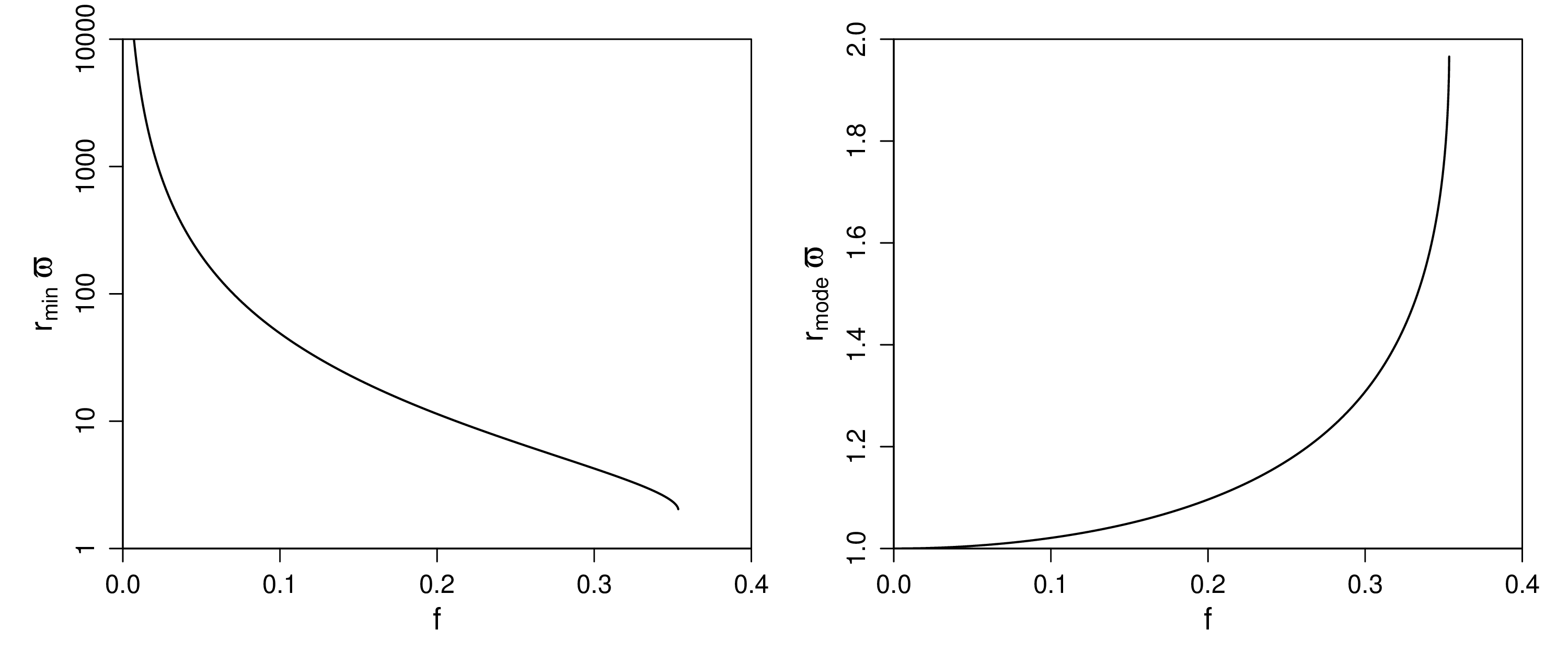}
\caption{The value of the roots (equation \ref{eqn:r2post_roots}) of the posterior $\upost[r^2]$ (equation~\ref{eqn:upostr2}, which uses
the constant volume density prior) as a function of $\fpe$. The two roots are the minimum ($\rmin$, left) and the mode ($\rmode$, right). $\parallax$ is the measured parallax. \label{fig:roots_deriv_d.post_r2Prior}}
\end{center}
\end{figure}

We can find the turning points of the posterior by solving $d\upost[r^2]/dr=0$ for $r$. This gives a quadratic equation with solutions
\begin{equation}
\rmin, \rmode \,=\, \frac{1}{\parallax}\frac{1}{4\fpe^2}\left(1 \pm \sqrt{1-8\fpe^2} \right) \ .
\label{eqn:r2post_roots}
\end{equation}
The solution with the positive sign is the minmum ($\rmin$) and that with the negative sign is the mode ($\rmode$).
They are plotted as a function of $\fpe$ in Figure \ref{fig:roots_deriv_d.post_r2Prior}.
There is an additional mode at $r=\rlim$ due to the discontinuous cut-off in the prior.
We see that for $\fpe$\,$>$\,$1/\sqrt{8}$ (=\,0.354) neither turning point is defined: the posterior transitions from having two modes and a minimum to the single mode at $r=\rlim$. Thus for larger values of $\fpe$ we would need to resort to another estimator; $\rlim$ itself would be consistent.

Equation \ref{eqn:r2post_roots} shows that the mode scales linearly with $1/\parallax$:
the function in the right panel of Figure \ref{fig:roots_deriv_d.post_r2Prior} is the factor which we multiply $1/\parallax$ by to estimate the distance. It depends on the parallax error (through $\fpe$), but not on $\rlim$.

Contrary to expectations (perhaps), the posterior would have a minimum for all (positive) values of $\fpe<1/\sqrt{8}$, but if it occurs at $r>\rlim$ it will not be seen due to the cut off from the prior. In the case shown $\rlim \parallax = 10$, so the minimum will only drop below $\rlim$ when $f>0.212$.\footnote{This comes from solving the positive sign solution of equation \ref{eqn:r2post_roots} for $\fpe$ ($>0$) with $\rmin=\rlim$. The solution is
\begin{equation}
\fpe \,=\, \sqrt{\frac{1}{2\rlim\parallax}\left(1 - \frac{1}{\rlim\parallax}\right)} \ .
\end{equation}
}
The minimum goes to infinity in the limit $\fpe \rightarrow 0$.

\begin{figure}
\begin{center}
\includegraphics[width=0.49\textwidth, angle=0]{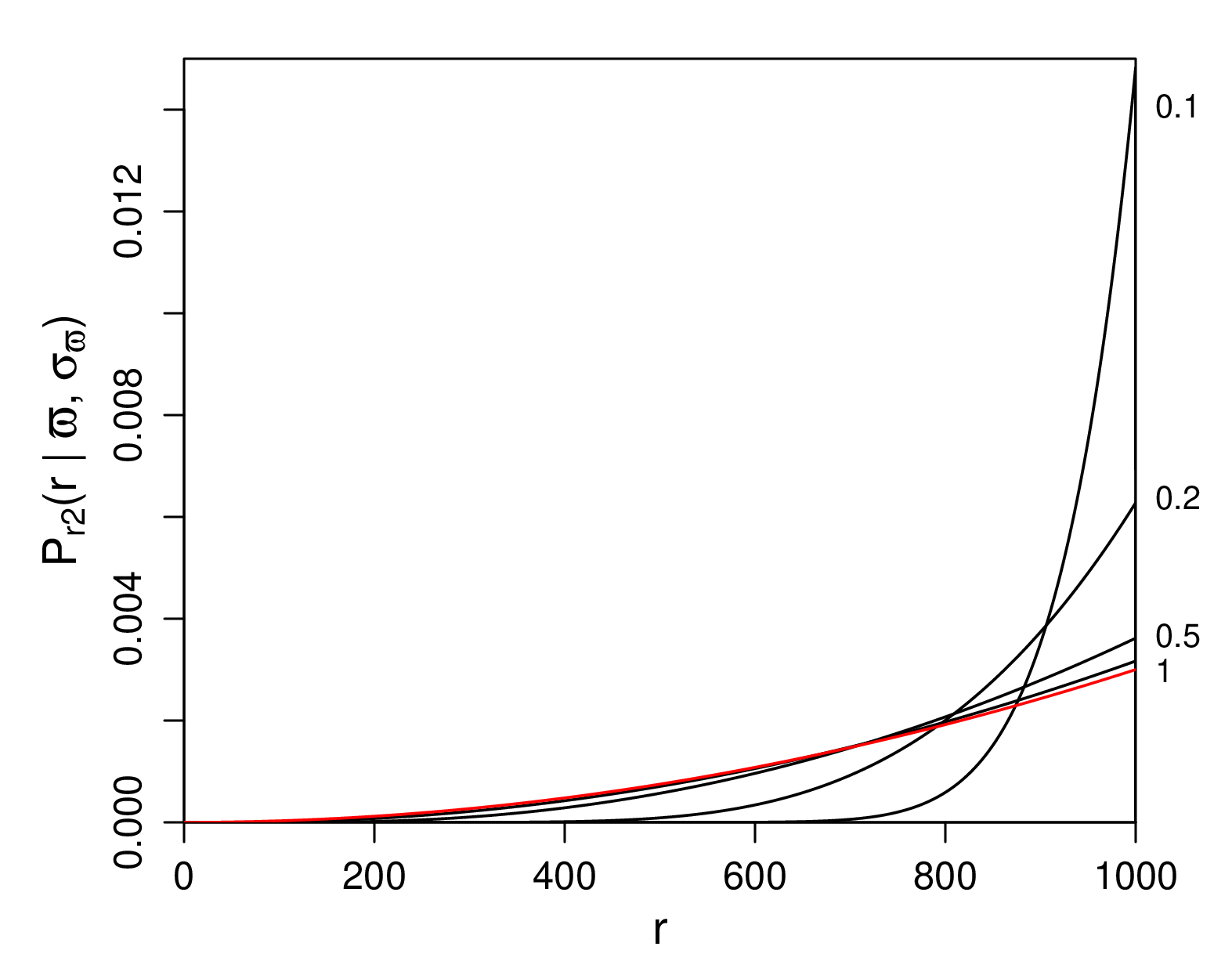}
\caption{The normalized posterior $\post[r^2]$ (truncated constant volume density prior with $\rlim=10^3$)
for $\parallax=-1/100$ and four values of $|\fpe|=(0.1, 0.2, 0.5, 1.0)$ (black lines). The red line shows the posterior for $\parallax=0$ for $\parsd \gg 1/\rlim$ ($\fpe$ is then undefined).\label{fig:post_r2Prior_nonposw}}
\end{center}
\end{figure}

The posterior is also defined when $\parallax \leq 0$, as can be seen by inspection of equation \ref{eqn:upostr2} and shown in Figure \ref{fig:post_r2Prior_nonposw}. This demonstrates that when we use a prior, negative or zero parallaxes provide information in agreement with our intuition: they are just noisy measurements which happened to end up non-positive. They imply the distance is likely to be large (up to the constraint imposed by prior knowledge), and the range of probable solutions depends on the size of the parallax error (a smaller error means more constraint). This intuition is expressed quantitatively by the probabilistic approach.

Let us now test this estimator with the same empirical test as used before (section \ref{sec:empirical_test}). I draw the data from the constant volume density prior (equation \ref{eqn:r2prior}) with the same value of $\rlim$ as used in the posterior ($10^3$). My distance estimator is the lower mode of the posterior when it is defined, i.e.\ $\rmode$ in equation \ref{eqn:r2post_roots} if $\fpe<1/\sqrt{8}$ and $\parallax>0$, and $\rlim$ otherwise.
Note that this decision can be made using only measured data (it involves $\fpe$, not $\fpetrue$).
If $\rmode>\rlim$ -- an extreme mode -- then I set $\rest=\rlim$, because this is what the prior tells us. Thus
\begin{equation}
  \rest  \,=\,  \begin{dcases}
  \rmode  & \:{\rm if}~~ \parallax > 0  \ \ {\rm and} \ \ \fpe < 1/\sqrt{8} \ \ {\rm and} \ \ \rmode \leq \rlim \\
  \rlim      & \:{\rm if}~~ \parallax > 0  \ \ {\rm and} \ \ \fpe < 1/\sqrt{8} \ \ {\rm and} \ \ \rmode > \rlim \hspace*{1em}{\rm (extreme~mode)} \\
  \rlim      & \:{\rm if}~~ \parallax > 0  \ \ {\rm and} \ \ \fpe \geq 1/\sqrt{8} \hspace*{1em}{\rm (undefined~mode)} \\
  \rlim      & \:{\rm if}~~ \parallax \leq 0 \\
\end{dcases}
\label{eqn:postr2t_mode}
\end{equation}

\begin{figure}
\begin{center}
\includegraphics[width=0.49\textwidth, angle=0]{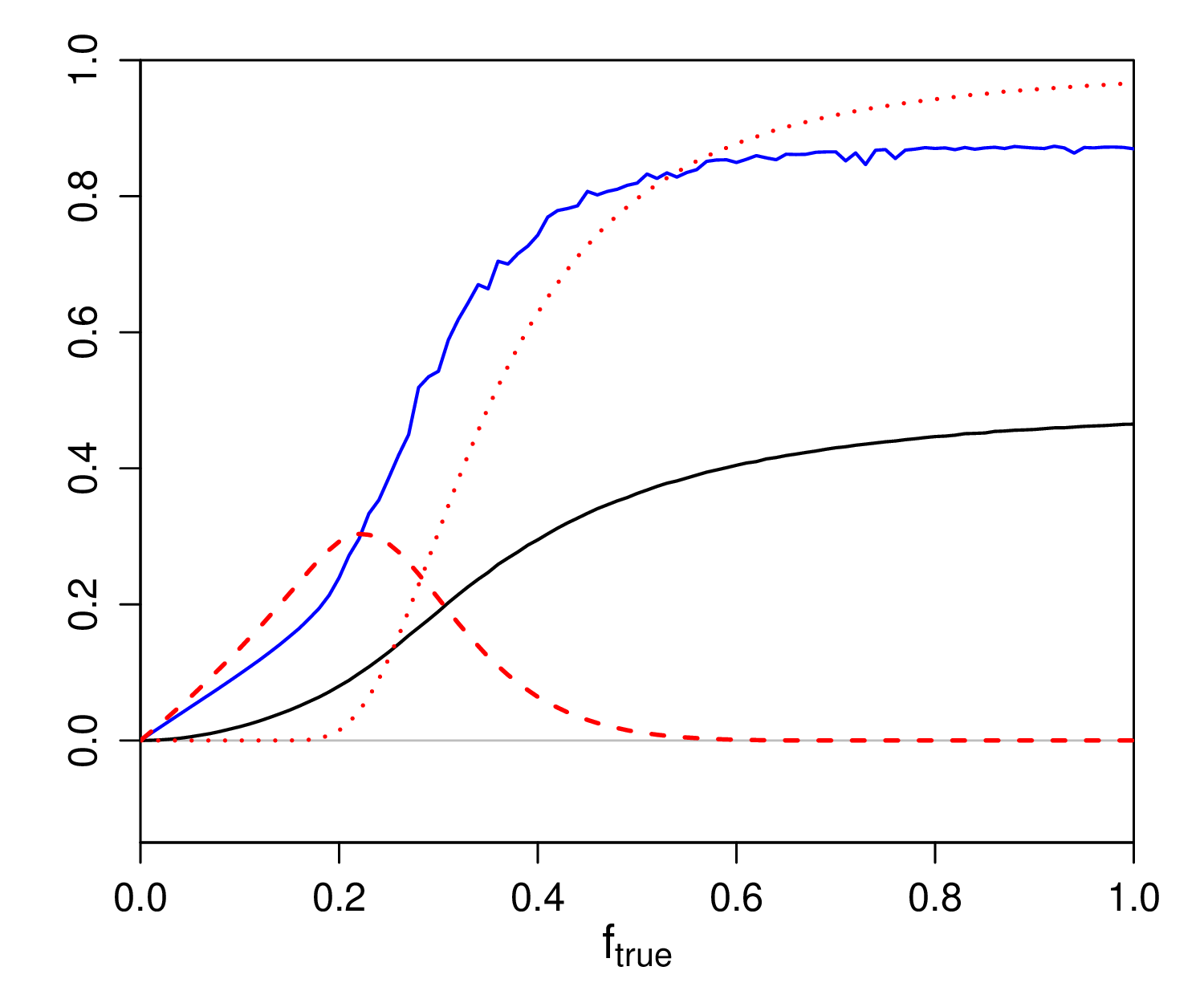}
\includegraphics[width=0.49\textwidth, angle=0]{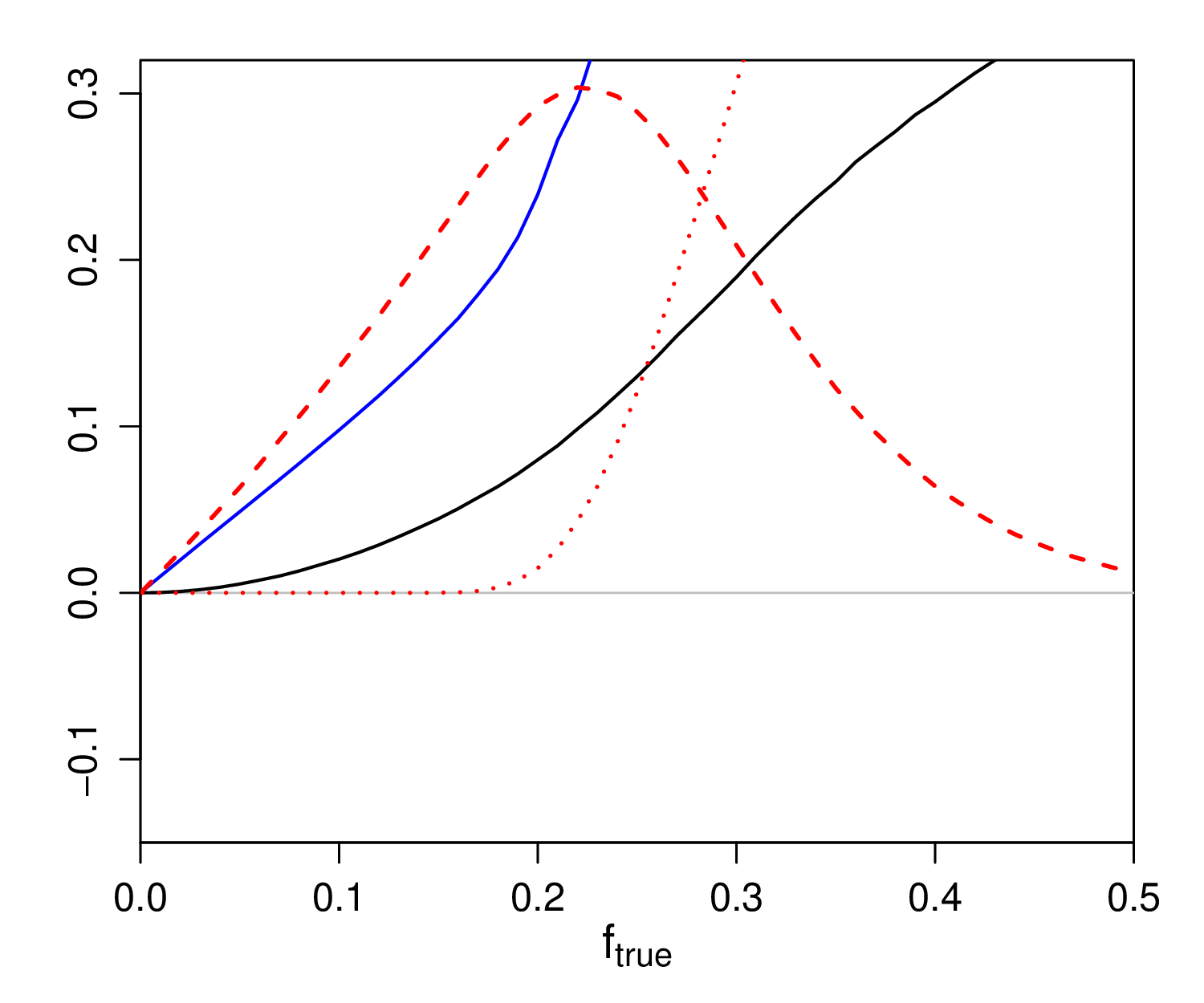}
\caption{
The bias (black) and standard deviation (blue) as a function of $\fpetrue$ for
the mode estimator (equation \ref{eqn:postr2t_mode}) of the posterior $\upost[r^2]$ (constant volume density prior;
equations \ref{eqn:upostr2} and \ref{eqn:r2post_roots}) for data drawn from the same prior. The right panel is a zoom of the left panel. 
The dashed red line shows the fraction of samples with an extreme mode ($\rmode > \rlim$).
The dotted red line
shows the fraction of samples with an undefined mode ($\fpe\geq1/\sqrt{8}$). 
In both these cases the distance estimates for the samples are set to $\rlim$.
\label{fig:scaledResiduals_mode_r2Prior_rmax1e3_truncateMode}}
\end{center}
\end{figure}

The results are shown in Figure \ref{fig:scaledResiduals_mode_r2Prior_rmax1e3_truncateMode}. 
Note that the curve is defined for all positive value of $\fpetrue$ (the horizontal axis), in particular beyond $1/\sqrt{8}$.
At any given value of $\fpetrue$ the samples will have a range of values of $\fpe$
due to the parallax sampling in step \ref{listitem:draw_parallax} of the empirical test. But 
these still contribute to the sample for the specific value of $\fpetrue$.

We see in the figure how the fraction of samples with an undefined mode (dotted red line) increases steadily with $\fpetrue$. The majority of the samples fall in this category once $\fpetrue>0.35$. This largely explains why the bias and standard deviation saturate: setting most distances to $\rlim$ limits the distance error incurred. The fraction of samples with an extreme mode (dashed red line) initially increases, but then decreases, because for large parallax errors the measured parallax is increasingly likely either to be negative (which is not considered an extreme mode) or positive but so small that $\fpe \geq 1/\sqrt{8}$, in which case the mode is undefined and the sample contributes to the dotted curve instead.

\begin{figure}
\begin{center}
\includegraphics[width=0.49\textwidth, angle=0]{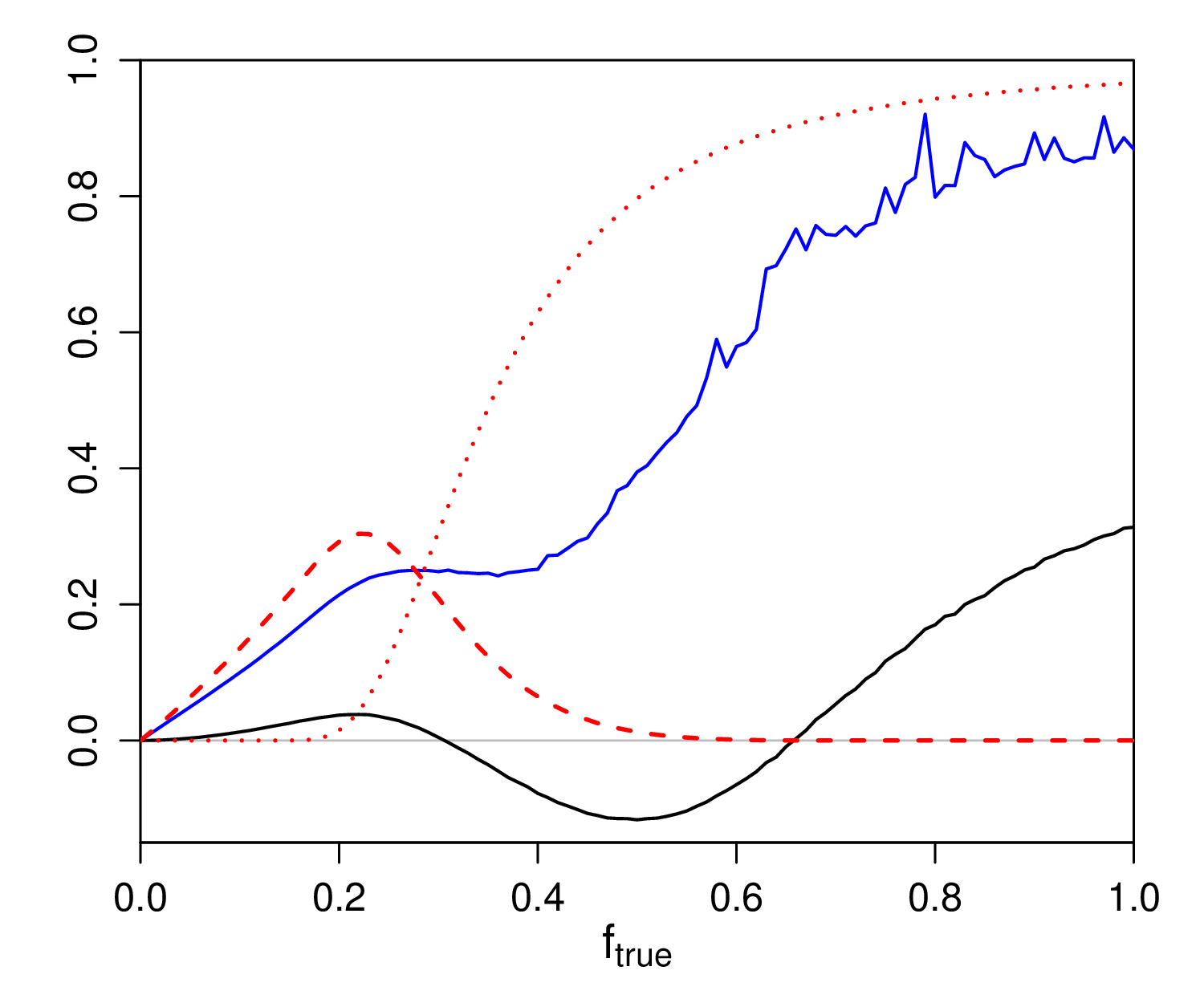}
\includegraphics[width=0.49\textwidth, angle=0]{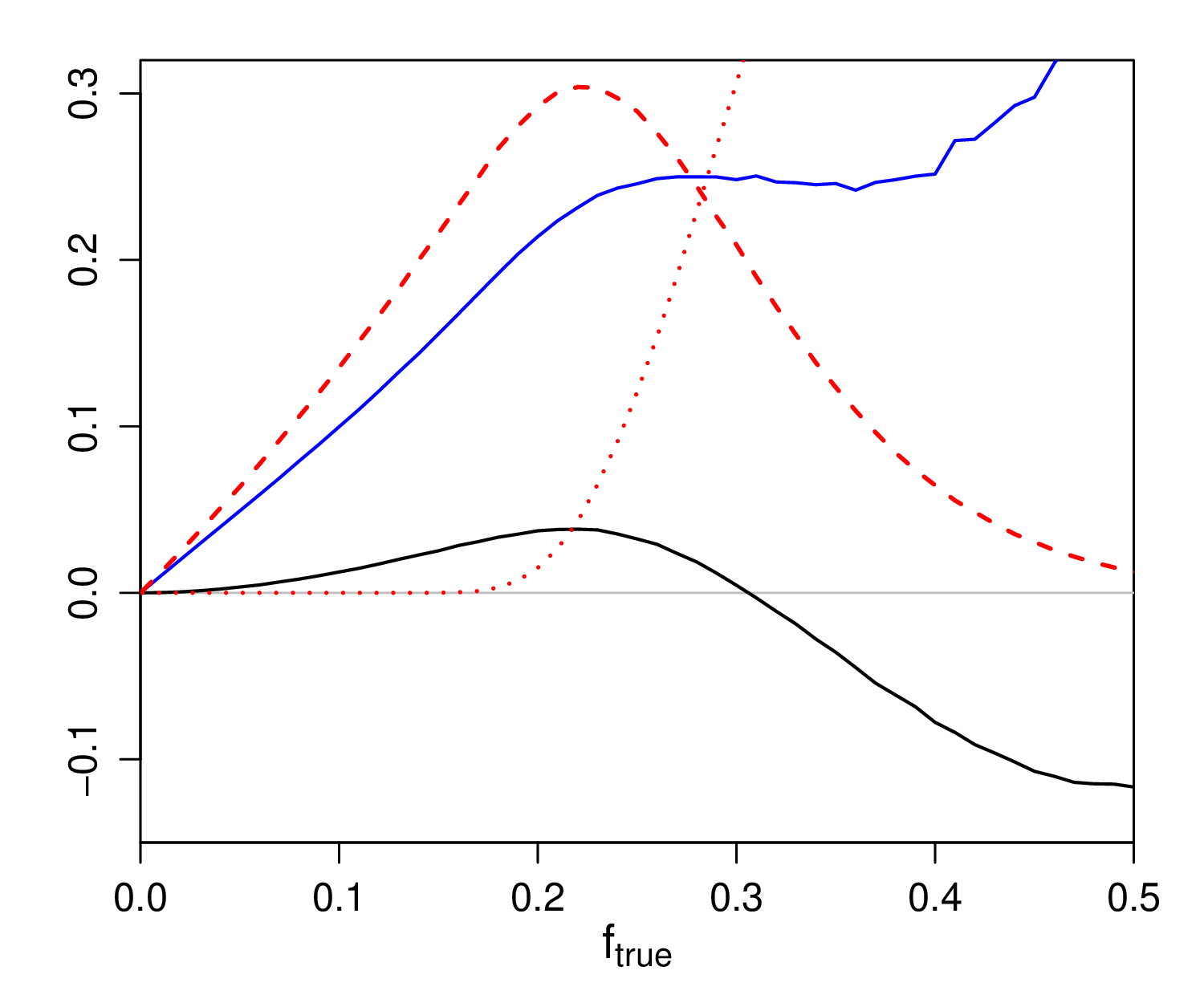}
\caption{
As Figure \ref{fig:scaledResiduals_mode_r2Prior_rmax1e3_truncateMode}, but now the samples with extreme and undefined modes are rejected rather than set to $\rlim$.
\label{fig:scaledResiduals_mode_r2Prior_rmax1e3_rejectMode}}
\end{center}
\end{figure}
 
Using this prior and strategy, we get much better results than with the improper uniform prior (Figure \ref{fig:scaledResiduals_mode_runifPrior_rmax1e3}). We also achieve a lower standard deviation than with the truncated uniform prior (Figure \ref{fig:scaledResiduals_mode_runifPrior_rmax1e3_truncateMode}) for larger values of $\fpetrue$, but at the price of a much larger bias. 
Note that these results are for data drawn from the same prior as that used in the posterior. With a mismatched prior we will get different results.

Instead of truncating the extreme and undefined mode values to $\rlim$, we could just reject them, as we did with the truncated uniform prior in Figure \ref{fig:scaledResiduals_mode_runifPrior_rmax1e3_rejectMode}. The results of this are shown in Figure \ref{fig:scaledResiduals_mode_r2Prior_rmax1e3_rejectMode}. The dotted and dashed lines are the same (they are statistically the same data), but now the standard deviation and especially the bias look quite different. Indeed, the bias now turns negative for some values of $\fpetrue$. This is because we are rejecting exclusively samples which would otherwise be assigned the largest distance possible. This also helps to lower the standard deviation. But once $\fpetrue$ increases nearly to 1 (100\% expected parallax errors), the standard deviation and bias are large again. Even worse, we have had to reject over 95\% of the data.

While a constant volume density prior is more desirable from a physical point of view than the prior uniform in $r$, it suffers from the same problem: a discontinuous cut-off in $\rest$. For lower accuracy parallaxes, this demands that we either throw away those data or assign a fixed distance, $\rlim$. Both are undesirable.

\section{An exponentially decreasing volume density prior}\label{sec:expdec}

\begin{figure}
\begin{center}
\includegraphics[width=0.49\textwidth, angle=0]{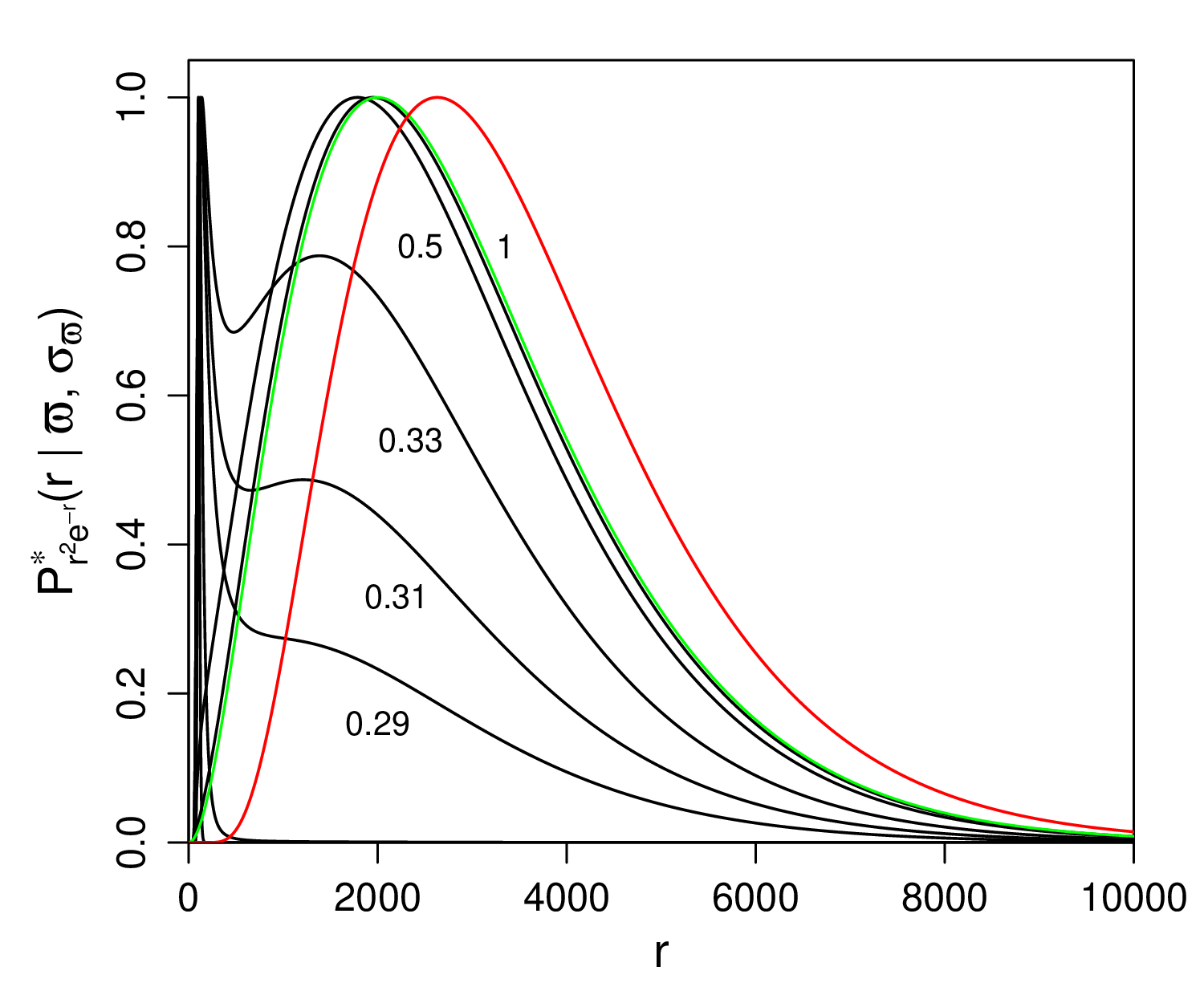}
\includegraphics[width=0.49\textwidth, angle=0]{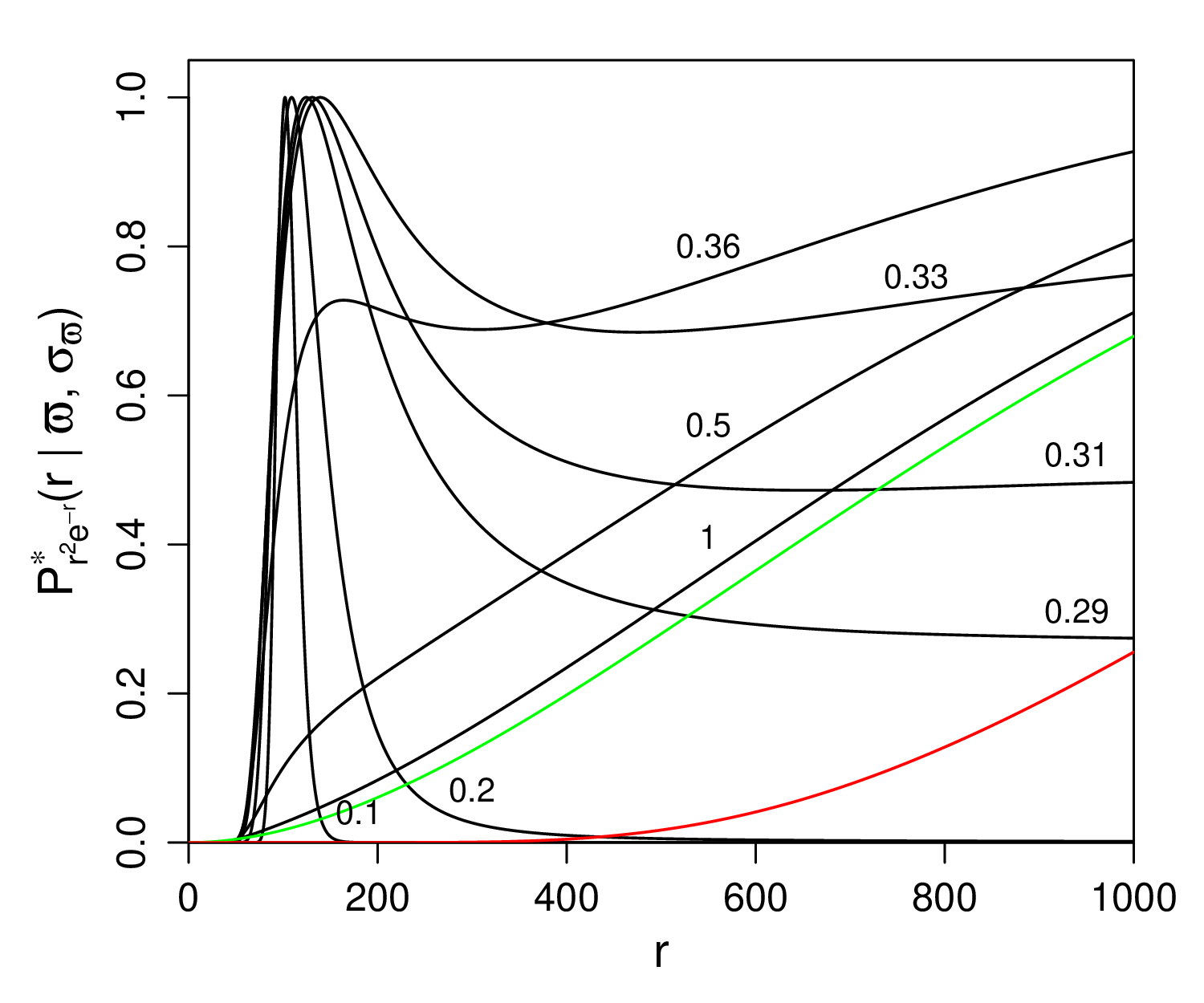}
\caption{The black lines in the left panel show the unnormalized posterior $\upost[r^2e^{-r}]$ (exponentially decreasing volume density prior; equation~\ref{eqn:upostr2e-2}) for $\rlen=10^3$, $\parallax=1/100$ and seven values of $\fpe=(0.1, 0.2, 0.29, 0.31, 0.33, 0.5, 1.0)$. 
The red line is the posterior for $\parallax = -1/100$ and $|\fpe|=0.25$. The green curve is the prior.
The right panel is a zoom of the left one and also shows an additional posterior for $\fpe=0.36$.
All curves have been scaled to have their highest mode at $\upost[r^2e^{-r}]=1$ (outside the range for some curves in the right panel). \label{fig:ud.post_r2e-rPrior}}
\end{center}
\end{figure}

We would like to have a posterior which yields a distance estimate for any value of $\fpe$ and $\parallax$, including non-positive parallaxes. This can be achieved by replacing the sharp cut-off of the previous priors with something which drops asymptotically to zero as $r \rightarrow \infty$.
Here I investigate a prior which produces an exponential decrease in the volume density of stars, $P(V) \sim \exp(-r/\rlen)$, namely
\begin{equation}
\prior[r^2e^{-r}]  \,=\,  \begin{dcases}
  \frac{1}{2\rlen^3}\,r^2e^{-r/\rlen}  & \:{\rm if}~~ r >0 \\
  0                          & \:{\rm otherwise}
\end{dcases}
\label{eqn:r2e-2prior}
\end{equation}
where $\rlen>0$ is a length scale. The unnormalized posterior is
\begin{equation}
\upost[r^2e^{-r}]  \,=\,  \begin{dcases}
  \frac{r^2e^{-r/\rlen}}{\parsd} \exp{ \left[ -\frac{1}{2\parsd^2}\left(\parallax-\frac{1}{r}\right)^2 \right] }  & \:{\rm if}~~ r > 0 \\
  0                          & \:{\rm otherwise} \ .
\end{dcases}
\label{eqn:upostr2e-2}
\end{equation}
Examples of this posterior are shown in Figure
\ref{fig:ud.post_r2e-rPrior}. We see a dependence on $\fpe$ which is similar to that of the $r^2$ prior for small $r$, but now with a smooth peak at large $r$ and a continuous decline beyond. 
Depending on the value of $\fpe$, we may have one or two modes. In this example there is a single mode for 
$0 < \fpe \lesssim 0.30$, two modes for $0.30 \lesssim \fpe \lesssim 0.373$, and one mode for $\fpe \gtrsim 0.373$.
Note that although $\rlen$ is a characteristic length scale of the posterior, the mode corresponding to this is at a somewhat larger value of $r$. 

The larger $\fpe$, the less informative the data, and in the limit $\fpe \rightarrow \infty$ 
the posterior just becomes the prior for any value of the parallax (positive, zero, or negative). Indeed, already at $\fpe=1$ the posterior is almost indistinguishable from the prior (the green line). 

The red line in Figure \ref{fig:ud.post_r2e-rPrior} shows the posterior for a negative parallax of $-1/100$ with $|\fpe|=0.25$. If $|\fpe|$ were smaller this posterior would shift to the right. This makes sense, because a smaller $|\fpe|$ means we are more confident that the true parallax is close to zero. As $|\fpe|$ gets larger the posterior shifts to the left, eventually converging to the prior.

To find the mode we set $d\upost[r^2e^{-r}]/dr=0$ which gives
\begin{equation}
\frac{r^3}{\rlen} - 2r^2 + \frac{\parallax}{\parsd^2}r - \frac{1}{\parsd^2} \,=\, 0 \ .
\label{eqn:r2e-rpost_roots}
\end{equation}
This is a cubic equation which generally has three complex roots, but fewer solutions here due to the physical limitations on the signs of the variables.  The roots are a function not only of $\fpe = \parsd/\parallax$, but also of $\rlen$ and $\parsd$. Depending on these values, the posterior may have three real roots, corresponding to two modes and one minimum, or just one real root, corresponding to a single mode.

Inspection of the roots leads to the following strategy for assigning the distance estimator, $\rmode$, from the modes:
\vspace*{-0.5em}
\begin{myitemize}
\item If there is one real root, it is a maximum: select this as the mode.
\item If there are three real roots, there are two maxima:
\begin{myitemize}
\item If $\parallax \geq 0$, select the smallest root (value of $r$) as the mode.
\item If $\parallax < 0$, select the mode with $r>0$ (there is only one).
\end{myitemize}
\end{myitemize}
\vspace*{-0.5em}
The other two possibilities (zero or two real roots) do not occur for $\parsd>0, \rlen>0$.  

\begin{figure}
\begin{center}
\includegraphics[width=0.98\textwidth, angle=0]{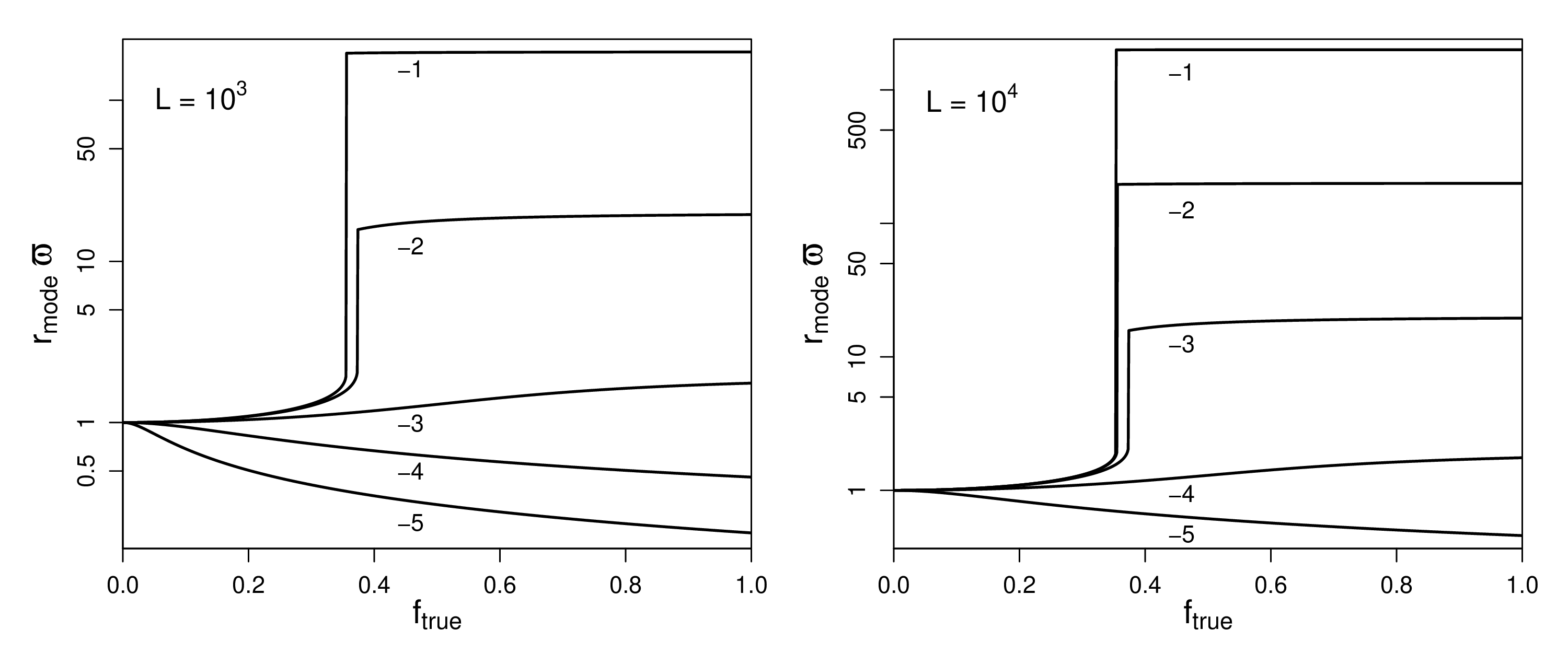}
\caption{The distance estimator mode, $\rmode$, of the posterior $\upost[r^2e^{-r}]$ (which uses the exponentially decreasing volume density prior; equation~\ref{eqn:upostr2e-2}) shown as $\rmode \parallax$ as a function of $\fpetrue$. Each line is for a different value of the parallax, $\parallax$, and labelled with $\log_{10}{\parallax}$.
The left panel is for $\rlen=10^3$ and the right panel for $\rlen=10^4$. Note the (different) log scales on the vertical axes. \label{fig:roots_mainMode_d.post_r2e-rPrior}}
\end{center}
\end{figure}
The variation of $\rmode$ as a function of $\fpe$ for different $\parallax$ and $\rlen$ is shown in Figure \ref{fig:roots_mainMode_d.post_r2e-rPrior}.\footnote{For a given $\rlen$ and $\parallax$, the variation of $\rmode\,\parallax$ with respect to $\fpetrue$ is independent of $\parallax$.}  Let us follow the curve labelled ``$-2$'' ($\parallax = 1/100$) in the left panel, which corresponds to the posteriors plotted in Figure~\ref{fig:ud.post_r2e-rPrior}.  At small values of $\fpetrue$ (below 0.30), the posterior has a single mode, and the value of $\rmode$ (for a given $\parallax$) increases slowly and smoothly with increasing fractional parallax error.  As $\fpetrue$ rises above about 0.30, a second mode appears, but I continue to use the mode at the lower value of $r$ as the distance estimator, because we can think of this one as being dominated by the data: it continues to evolve smoothly from the data-dominated regime (small $\fpetrue$). Once $\fpetrue$ rises above 0.373, there is a sudden increase in the value of $\rmode$. This corresponds to the ``data-dominated'' mode disappearing, leaving only the other,``prior-dominated'', mode for all larger values of $\fpetrue$.

If the measured parallax were smaller, say $\log_{10}\parallax =-3$, but $\rlen$ the same, then we see from Figure \ref{fig:roots_mainMode_d.post_r2e-rPrior} (left panel) that the posterior only ever has one mode for all $\fpetrue$. This is because the data and the prior are now indicating distances on a similar order of magnitude. If we instead used a larger $\rlen$ for $\log_{10}\parallax =-3$ (right panel, line labelled ``$-3$''), the measured parallax is again quite different from the prior, so we again see two modes for smaller $\fpetrue$, and the transition to a single mode for larger $\fpetrue$.
We only and always get such transitions when $\parallax\rlen \gg 1$. 
Whenever we have two modes, it is always the smaller one which is dictated by the data, so we can always make the correct choice of distance estimator.

\begin{figure}
\begin{center}
\includegraphics[width=0.49\textwidth, angle=0]{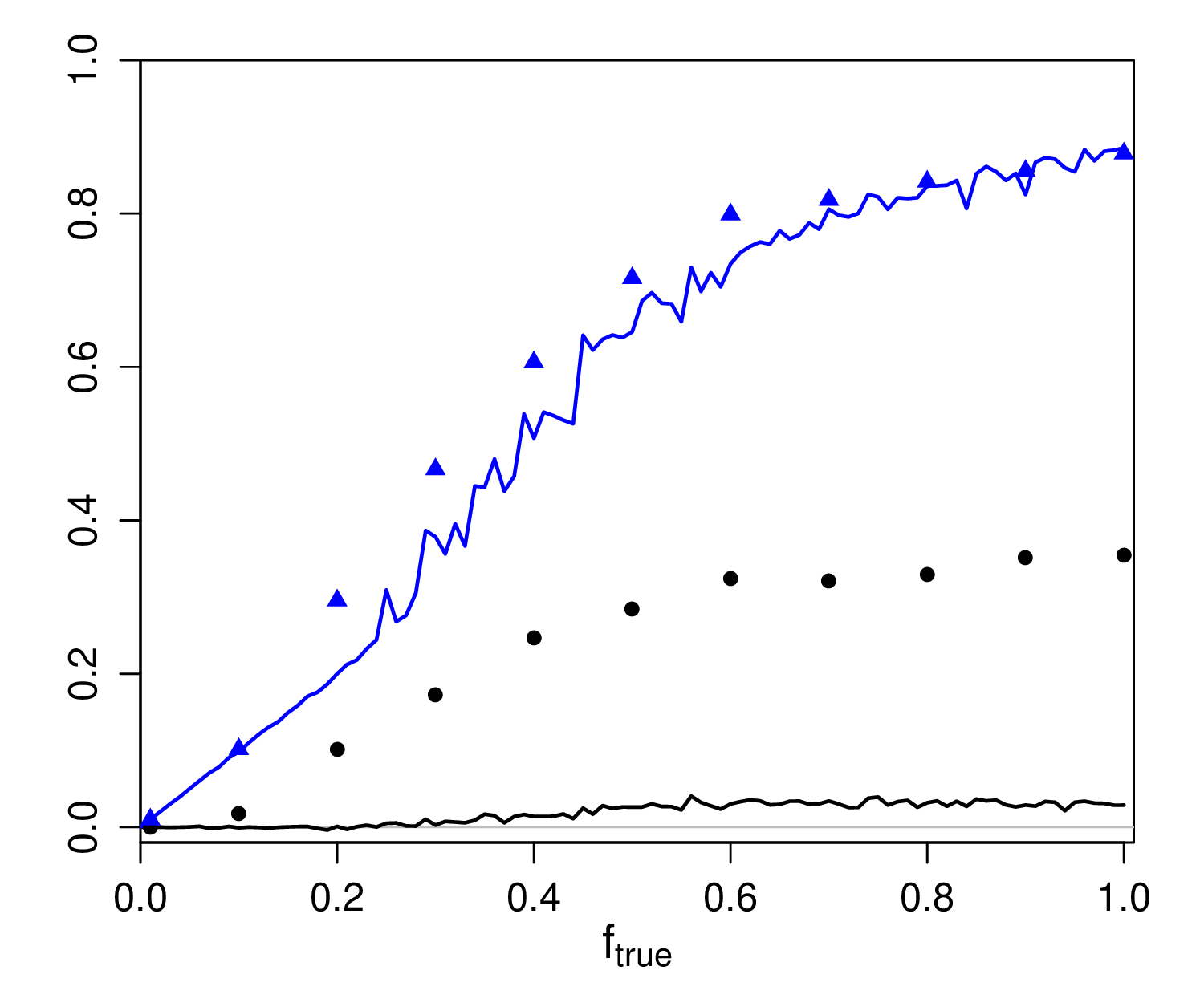}
\includegraphics[width=0.49\textwidth, angle=0]{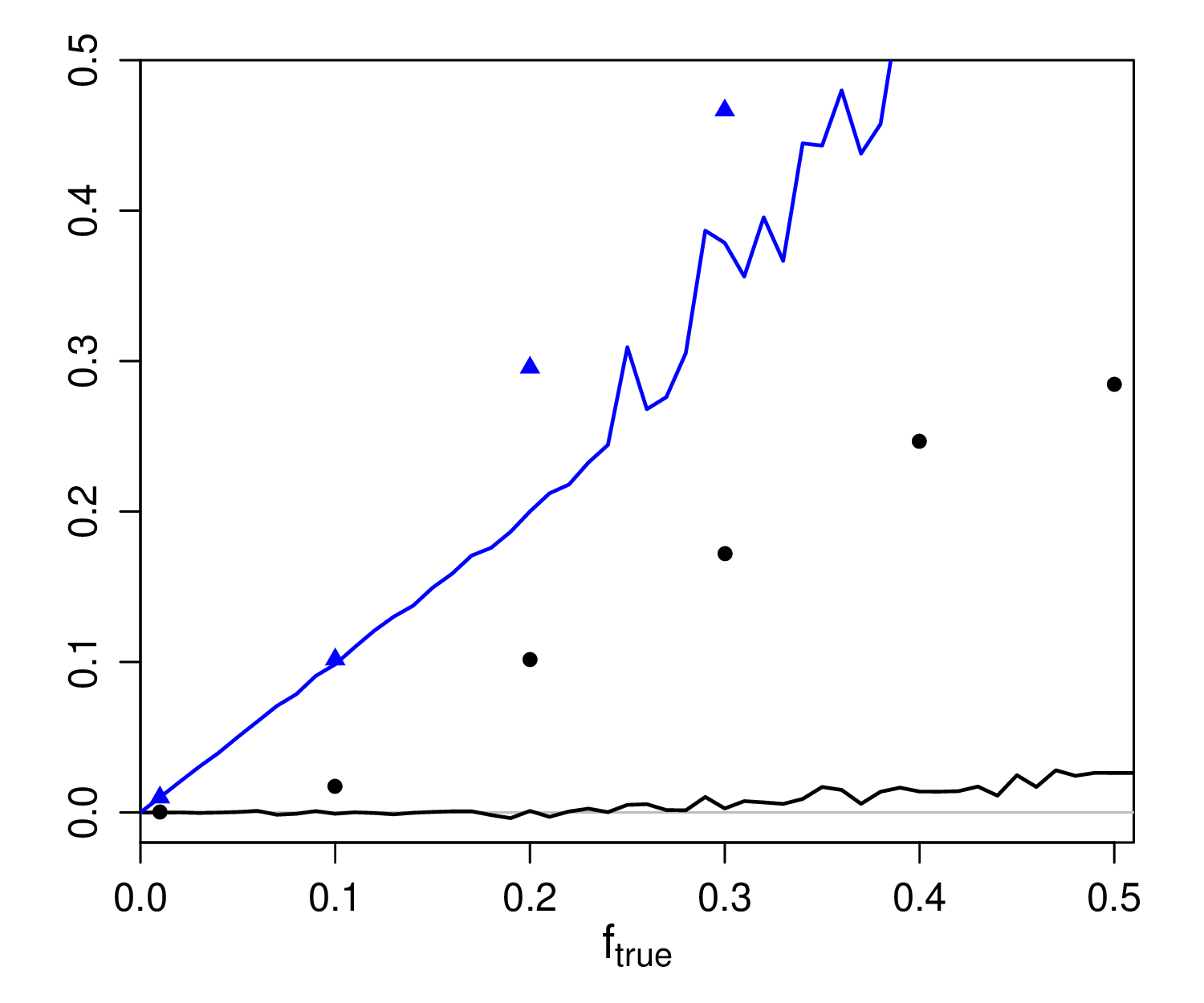}
\caption{The bias (black line) and standard deviation (blue line) as a function of $\fpetrue$ for
the mode distance estimator of the posterior $\upost[r^2e^{-r}]$ (exponentially decreasing volume density prior;  equation~\ref{eqn:upostr2e-2}) with $\rlen=10^3$ for data drawn from the same prior. 
The black circles and blue triangles are the bias and standard deviation respectively of the median of the posterior.
The right panel is a zoom of the left panel. \label{fig:scaledResiduals_mode_r2e-rPrior_rlen1e3}}
\end{center}
\end{figure}

To assess the performance of this posterior and estimator, I perform the empirical test drawing data from the same prior (equation~\ref{eqn:r2e-2prior}) as used in this posterior, both with $\rlen=10^3$. The results are shown in Figure \ref{fig:scaledResiduals_mode_r2e-rPrior_rlen1e3}.
The variation of the standard deviation is similar to what we have seen before with proper priors, and similar in scale to that obtained with the $r^2$ prior. The standard deviation increases linearly when $\fpetrue<0.25$ with a gradient of 1.0, slightly better than the $r^2$ prior in Figure \ref{fig:scaledResiduals_mode_runifPrior_rmax1e3_rejectMode}, and now without having to reject any data. But in the present case we get essentially no bias even for large $\fpe$. 
This lack of bias was not seen with the previous priors, even when we had a match between the prior and the distribution of true distances. 

The explanation for this is that we are now using a prior which decreases continuously and asymptotically after some distance. The particular choice of $\exp(-r/\rlen)$ is not important to achieve this. The problem with using a sharp cut-off in a prior is that a star with a sufficiently large fractional parallax error will, before applying the cut-off, have a significant probability for distances greater than the cut-off distance. If the mode lies beyond this value, applying the cut-off will cause a ``build up'' of inferred distances at the cut-off, resulting in a strong bias. This can occur even for stars with very large parallaxes (small distances): not only stars with a true distance near to the cut-off distance are affected.
If the fractional parallax error is large enough, the sharp cut-off will always cause a problem, and
increasing the cut-off distance will not help. This shows that truncating the prior at the largest distance expected based on the observed magnitude will not avoid the bias.
One might argue that truncating the prior
at very large distances will have little impact in practice, but with an asymptotically decreasing prior this is
even unnecessary, because at $r \gg L$ there is vanishingly small probability anyway.

Returning to Figure \ref{fig:roots_mainMode_d.post_r2e-rPrior}, we may be tempted to conclude that using $\rlen=1/\parallax$ would be a good idea, because it would give $\rmode\,\parallax \simeq 1$. But the parallax is a noisy measurement, so $\parallax \neq 1/\rtrue$. We
are not aiming to get $1/\parallax$ as our estimator (we saw in earlier sections how bad it is). The empirical test confirms that using $\rlen=1/\parallax$ in the prior gives poor results.

So far I have only investigated the mode of this posterior. Finding the mode is generally easy, because it involves differentiation and solving a polynomial equation. But it is not necessarily the best estimator in terms of bias and variance. Finding the mean and median (or any other quantile) involves integrals (semi-infinite and finite) of the form $r^n \upost[r^2e^{-r}]$ over $r$ for $n=1$
(and $n=2$ for the standard deviation, if we wanted this as an error estimate). These integrals have no simple solution, so we must resort to a numerical approach. Where possible I use adaptive quadrature techniques.
For some combinations of parameters this approach can find no solution, in which case I use the Metropolis algorithm.\footnote{After some trial and error I found that reasonably good results could be obtained using a step size equal to the mode of the posterior and initialized at the mode, with a chain length of $10^5$ samples and a short burn-in. A longer chain would reduce the noise, but takes longer to compute.} This is slow compared to adaptive quadrature, so I have only computed the bias and standard deviation of the median at a few values of $\fpetrue$. These are plotted in Figure \ref{fig:scaledResiduals_mode_r2e-rPrior_rlen1e3} as black circles and blue triangles (respectively). The median has a significant bias for larger $\fpetrue$. The mean has a similar profile, but with slightly larger values of both bias and standard deviation. So of all the obvious estimators, the mode appears to be the best. 

\begin{figure}
\begin{center}
\includegraphics[width=0.49\textwidth, angle=0]{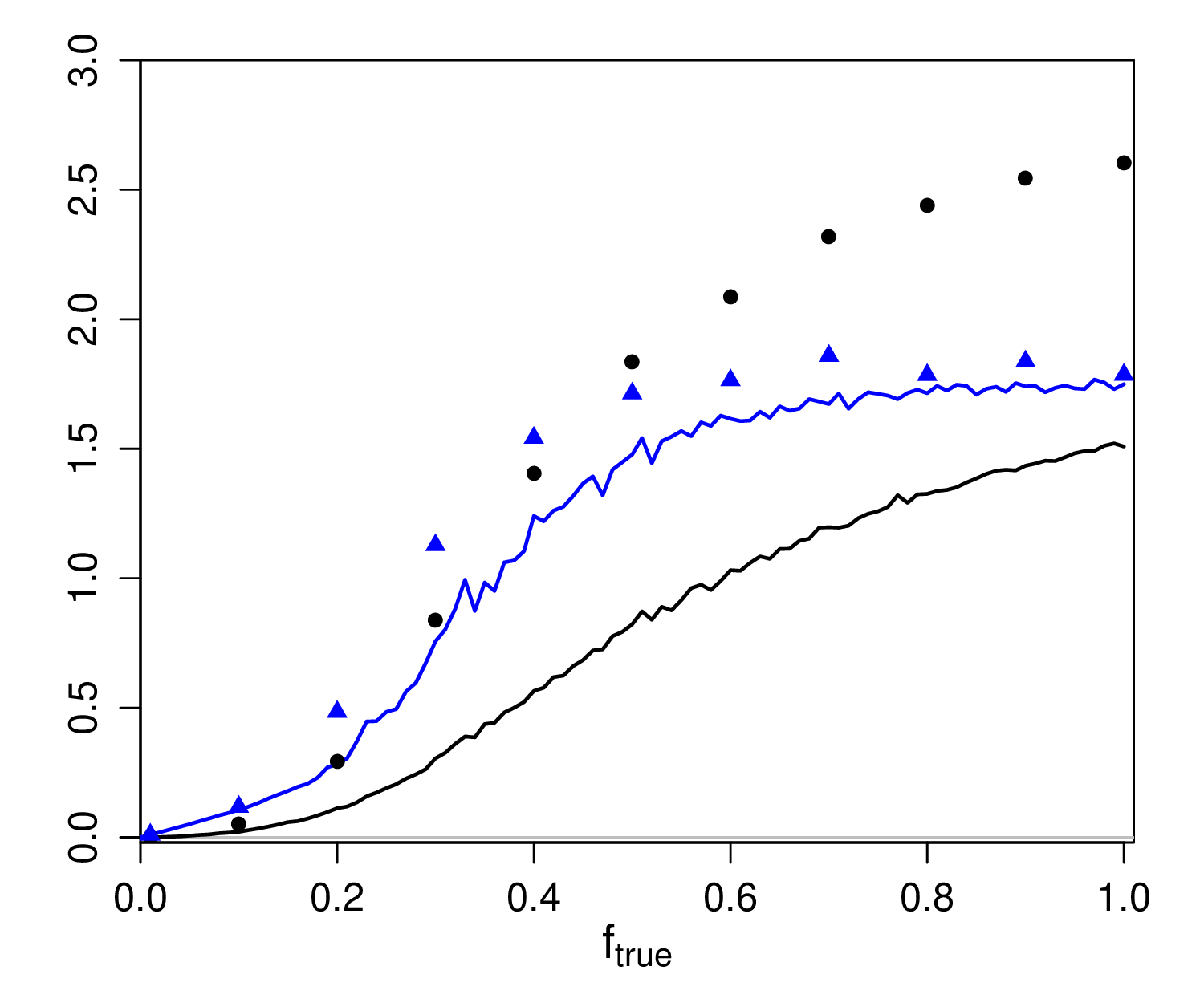}
\includegraphics[width=0.49\textwidth, angle=0]{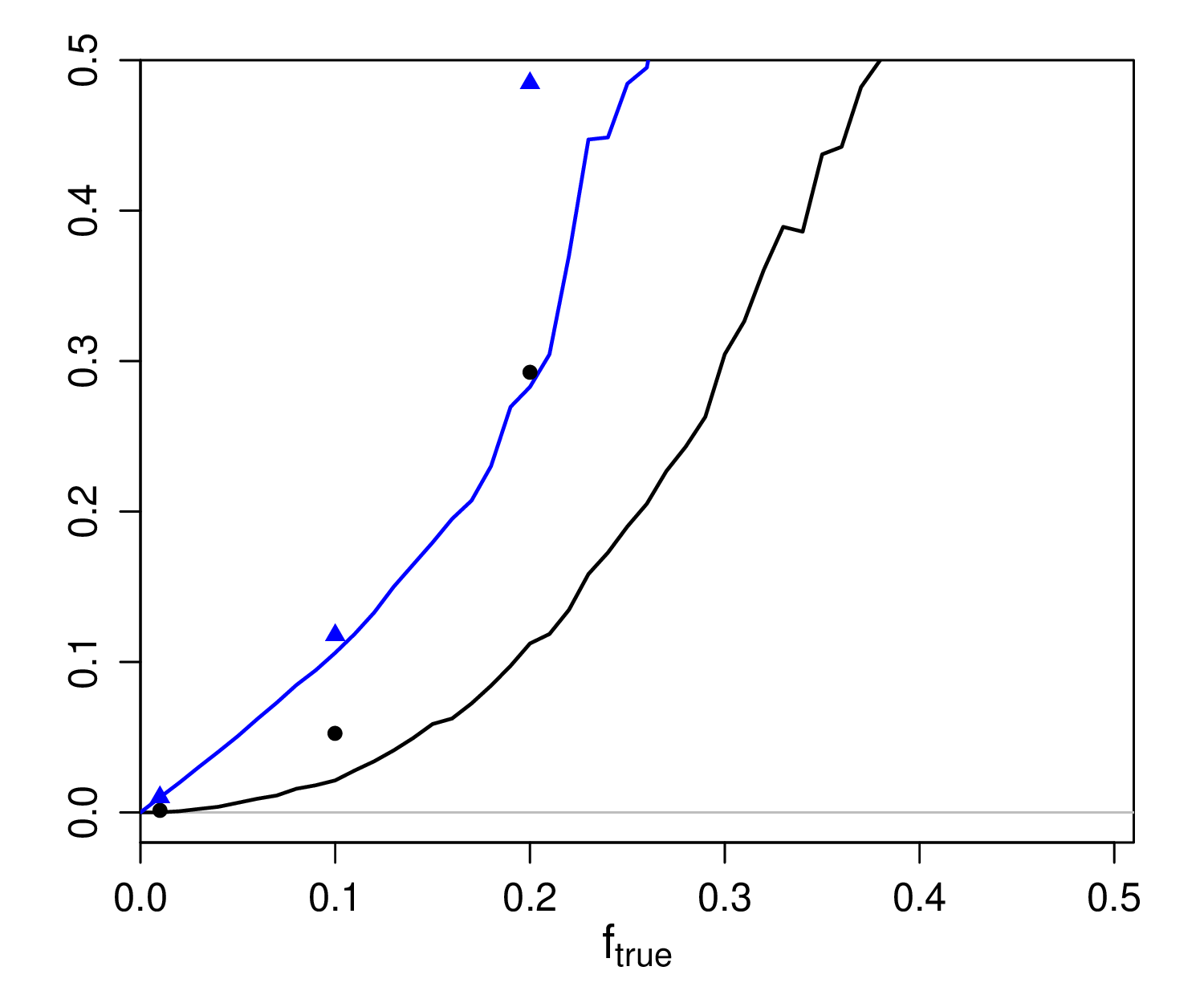}
\caption{As Figure \ref{fig:scaledResiduals_mode_r2e-rPrior_rlen1e3}, but now drawing data from the $r^2$ prior with $\rlim=10^3$ (and the left panel now has a different vertical scale).
\label{fig:scaledResiduals_mode_r2e-rPrior_rlen1e3_r2TruePrior_rmax1e3}}
\end{center}
\end{figure}

What happens if we have a mismatch between the distribution the data are drawn from, and the prior assumed in the model?  Figure \ref{fig:scaledResiduals_mode_r2e-rPrior_rlen1e3_r2TruePrior_rmax1e3} shows an example of this, where data are drawn from the truncated constant volume density prior ($P(r) \propto r^2$) with $\rlim=10^3$. As the posterior extends to considerably larger distances than the maximum {\em true} distance of any star in the sample, we inevitably overestimate distances when $\fpetrue$ is large: The estimation of course only sees the noisy measured parallax, which for large $\fpetrue$ has a high probability of being much smaller than $1/\rtrue$. If we repeat this experiment but now drawing data from a prior which extends to larger distances, e.g.\ $\rlim=10^4$, then the posterior estimator will tend to underestimate distances. There is not much more you can do about this other than to try to construct a well-matched prior {\em and} to report confidence intervals on your distance estimates by calculating quantiles on the posterior: I suggest 5\% and 95\%.  This interval will be large for large $\fpe$ (see the left panel of Figure \ref{fig:ud.post_r2e-rPrior}), but large fractional parallax errors necessarily mean that our distance estimates are poor and that we are dependent on the prior. Yet this is arguably preferable to throwing away data once $\fpe$ rises above some arbitrary value.

While this prior overcomes many of the problems we saw with other priors, I am not suggesting that we {\em always} use it. The point of the examples in the previous few sections was to show that the choice of prior (and estimator) can have a significant impact on the inferred distances, and that some of the obvious choices may be poor.  One should always think about what is an appropriate prior, given the available information. This is discussed further in section \ref{sec:choice}.  The main application of the exponentially decreasing volume density prior will be when we have very little information other than the parallax, and/or want to make minimal assumptions.

\section{A bias correction?}\label{sec:biascorrection}

Some of the literature on astrometry has been concerned with trying to apply bias corrections to using $1/\parallax$ as a distance estimator (e.g.\ Lutz \& Kelker 1973, Smith 2003, Francis 2013).  Unfortunately, many of these corrections are only valid under very restricted circumstances, only apply to small ($\lesssim 0.2$) fractional parallax errors, or are even wrong (Brown et al.\ 1997, Brown 2012).  Furthermore, we have just seen that there are priors which give unbiased estimates, so do we need to ever make bias corrections?  If there is a mismatch between the prior and the true distribution the stars are drawn from, then we might. But this requires that we know what the mismatch is, in which case we could improve the prior. It seems far better to always improve the prior than to make ad hoc corrections.

Suppose we did have a desirable estimator which is possibly biased (e.g.\ one which gives a lower standard deviation than an unbiased estimator).  Is there a useful correction which could be applied which does not simultaneously increase the standard deviation by too much?  This is an issue, because any bias correction is itself a function of measured and therefore noisy data.\footnote{Any ``correction'' which is independent of the data can be built into the model (e.g.\ the prior) so is no longer a correction.}  The bias and standard deviation plots I have shown, plotted against $\fpetrue$ ($= \parsd \rtrue$), show the bias and standard deviation we expect to obtain on average as a function of the expected fractional parallax accuracy.  They are useful to predict the quality of an estimator.  But they cannot be used to calculate a bias correction, because they depend on an unknown quantity. We would instead have to plot the bias against $\fpe$ ($ = \parsd/\parallax$). This could be done, but it is questionable whether the use of a suitable prior will ever bring up the need to make a bias correction.

\section{The choice of prior and estimator}\label{sec:choice}

There are many more priors and posterior-based estimators one could think of. 
A good prior is one which is as closely matched to the expected data as possible. 
In practice it should contain all the relevant information we have which is independent of individual measurements.
This could be a combination of both the expected intrinsic distribution of stars in the Galaxy,
how they enter the sample (as influenced by the choice of observational wavelength and interstellar extinction), and how they are selected by the survey (e.g.\ the survey detection efficiency and magnitude limit). This could be quite complex, and will generally vary with the direction of observation through the Galaxy. The exponentially decreasing volume density prior (section \ref{sec:expdec}) may be suitable when looking well out of the disk, where for a sufficiently deep survey the decrease in stellar density is caused mostly by the Galaxy itself rather than the survey. 
There are of course many variations on the theme of this prior. For example, we could produce a sharper decrease with increasing $r$ by using $\uprior = r^2\exp(-(r/L)^\alpha)$ for $\alpha > 1$.\footnote{In that case, the equation for the turning points is equation \ref{eqn:r2e-rpost_roots} but with $r^3/\rlen$ replaced by $(\alpha/\rlen^\alpha) r^{\alpha+2}$.}

Although stellar physics implies a maximum distance for any star, we have seen that discontinuities in the prior have the unpleasant property of leading to rejection of data or the cumulation of all poor data at the same maximum distance. 
For smooth priors, arbitrarily large distances receive asymptotically small probabilities, so are preferred.
If the survey includes extragalactic objects, one will want to allow for very large (but not infinite) distances.
This could be achieved using a two component prior (one for Galactic, one for extragalactic objects) with two modes.
When the posterior has two modes, one should generally report both, but may choose only to report one if it is significantly higher than the other.

Setting an informative prior which is wrong could be disastrous. When faced with a choice, I recommend using the simplest prior possible consistent with the constraints, as its influence on the results will be easier to interpret.  As with all cases of data analysis and interpretation -- whether Bayesian or not -- results will depend on personal choices.  A good approach is to test the sensitivity of the results to the use of different but equally acceptable priors. The poor quality data (high $\fpe$) are more affected by the choice of prior, and these will anyway have broader posteriors. Thus it is imperative that confidence intervals are reported on every distance estimate. I suggest to use 5\% and 95\% quantiles to define a 90\% confidence interval. The standard deviation, $\sigma$, should not be used, because $r$ is strictly positive. Quoting $r\pm\sigma$ implies a Gaussian and negative distances.

Given a posterior, one is still left with the choice of estimator. For the priors I have tested, the mode (taking care to deal with multimodality as described) is more accurate (lower bias and variance) and faster to calculate (no numerical integration) than the mean or median.

\section{Hipparcos and Gaia}

\begin{figure}
\begin{center}
\includegraphics[width=0.49\textwidth, angle=0]{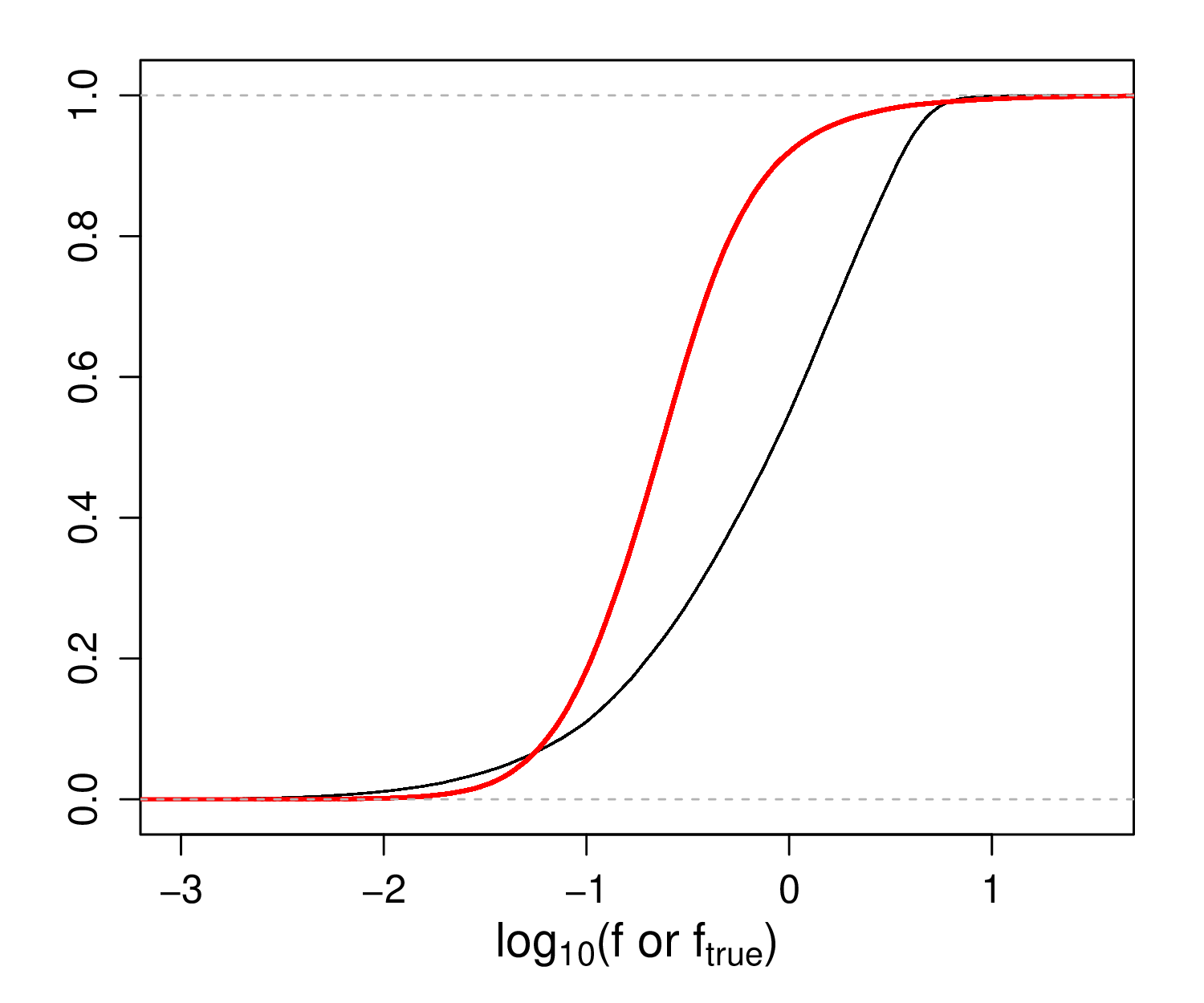}
\includegraphics[width=0.49\textwidth, angle=0]{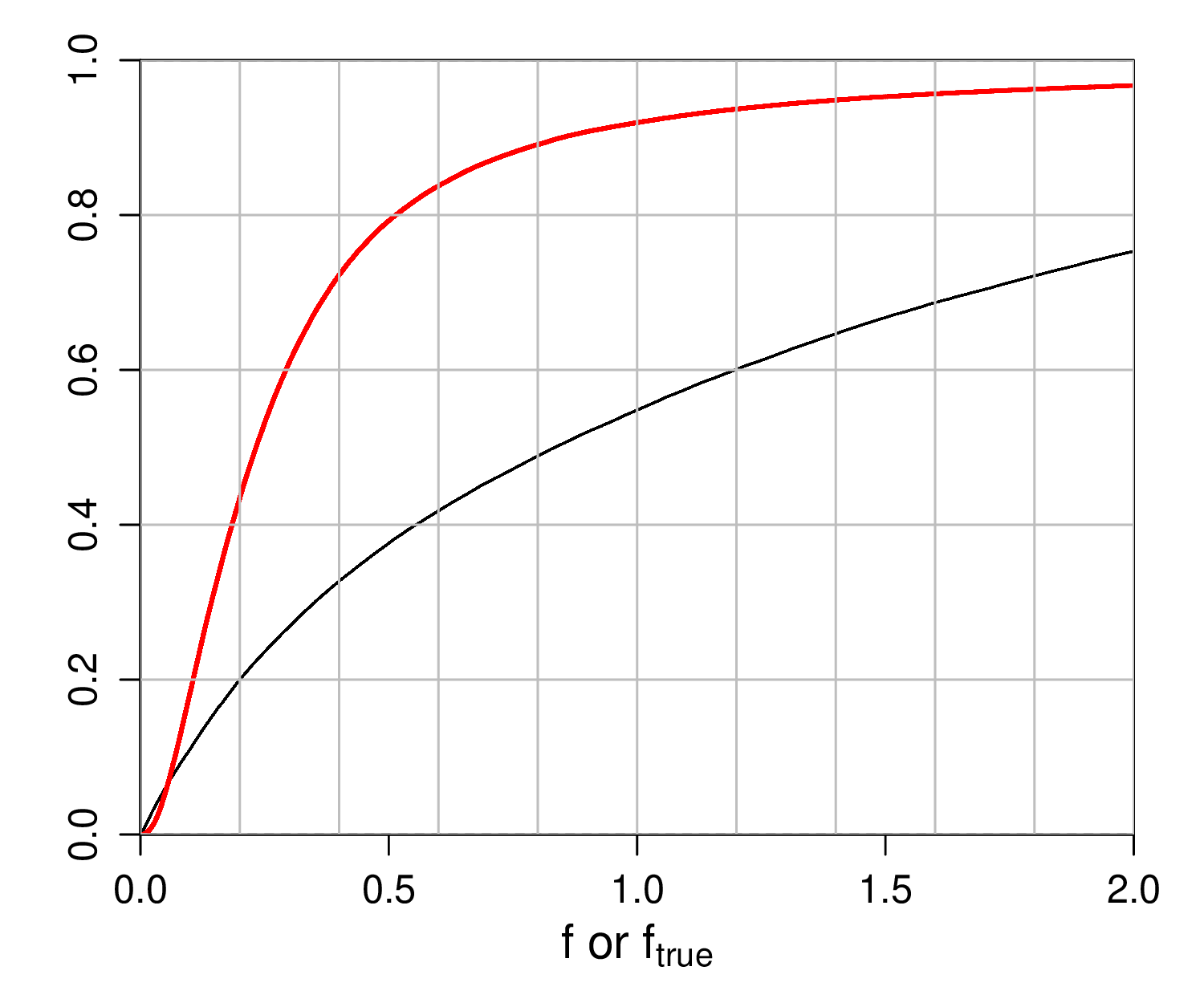}
\caption{Cumulative distribution of fractional parallaxes errors on a log scale (left) and a linear scale (right). The black line is the expected fractional parallax error (i.e.\ $\fpetrue$) for Gaia, calculated using the GUMS catalogue with the post-launch, sky-averaged, 5-year mission astrometric accuracy model. 
The red line is the measured fractional parallax error (i.e.\ $\fpe$) for the Hipparcos catalogue.
\label{fig:gums_hipparcos_cumf}}
\end{center}
\end{figure}

We have seen that when the fractional parallax errors are larger than about 20\%, use of a prior is imperative in order to make reasonable distance estimates at all and to assign confidence limits. Simply rejecting data with larger errors is in the best case a waste of hard-earned survey data, and in the worst case can severely bias analyses, because we will tend to reject fainter and/or more distant stars. While the accuracy of the ongoing Gaia survey will surpass all previous surveys in parallax accuracy on a large number of stars (accuracy of order 0.02\,mas at $15^{th}$ magnitude), it remains that many of the stars in the Gaia catalogue will have large fractional parallax errors. Using a published simulation of what Gaia is expected to observe (Robin et al.\ 2012) together with the first post-launch estimates of Gaia's end-of-mission, sky-averaged parallax accuracy (de Bruijne et al.\ 2015), I have calculated the fraction of stars with expected fractional parallax errors ($\fpetrue$) less than some amount: see the black line in Figure \ref{fig:gums_hipparcos_cumf}. Only about 20\% of the Gaia catalogue will have $\fpetrue<0.2$. This leaves at least 800 million stars requiring use of sensible prior information if we are to obtain a meaningful and not totally deviant distance estimate from the astrometry.  The situation is better with the Hipparcos catalogue (red line Figure \ref{fig:gums_hipparcos_cumf}), yet even here only about 40\% of the stars have $\fpe<0.2$.

For this reason, we should make use of astrophysical parameters which will be estimated from the Gaia spectrophotometry and spectroscopy (Bailer-Jones et al.\ 2013) to improve our distance estimates. Specifically, the interstellar extinction and absolute magnitude, combined with our knowledge of stellar astrophysics embodied in the Hertzsprung--Russell diagram, give an independent, probabilistic determination of the distance.  This posterior PDF can be combined with the astrometric-based posterior PDF to yield a distance estimate which is more accurate than either estimate alone.


\section{Summary and conclusions}

I have shown that estimating the distance given a parallax is not trivial. The naive approach of reporting $1/\parallax\,\pm\,\parsd/\parallax^2$ fails for non-positive parallaxes, is extremely noisy for fractional parallax errors larger than about 20\%, and gives an incorrect (symmetric) error estimate.  Probability-based inference for the general case is unavoidable.  Adopting an ``uninformative'' improper uniform prior over all positive $r$ does not solve any of these problems, and is in fact both informative and implausible (viz.\ a volume density of stars varying as $1/r^2$). The problems can be avoided by using a properly normalized prior which necessarily decreases after some distance. The use of a prior with a sharp cut-off is not recommended, because it introduces significant biases for low accuracy measurements at all distances (not just those near the cut-off).  Instead, a prior which converges asymptotically to zero as distance goes to infinity should be used.  Perhaps the simplest case is $\uprior = r^2\exp{(-r/\rlen)}$, which corresponds to an exponentially decreasing volume density with a length scale $\rlen$.  It is relatively neutral in that it is broad, decreases slowly, and has only one parameter. The mode of the corresponding posterior is an unbiased estimator (for the case that the population observed conforms to the prior) and it provides meaningful estimates for arbitrarily large parallax errors as well as non-positive parallaxes.  Bias corrections are not necessary.  The variance of this estimator behaves as well as one could expect given the irreducible measurement errors.  The median and mean perform less well than the mode.

Distance estimates must be accompanied by uncertainty estimates. For fractional parallax errors larger than 
about 20\%, the posterior over distance is significantly asymmetric, so we should always
report confidence intervals using quantiles (e.g.\ 5\% and 95\%) rather than standard deviations.  

The performance of any distance estimator depends, of course, on how well the prior is matched to the true distribution of distances. One must consider how much one is willing to tune the prior to previous knowledge of the Galaxy. In any case, priors with discontinuities should be avoided, and the sensitivity of results to the adopted priors should always be tested and reported.  Limiting inference to objects with fractional parallax errors much better than 20\% significantly diminishes the dependence of results on the choice of prior. But this would limit one to using only 20\% of the Gaia catalogue, which may introduce truncation biases into subsequent astrophysical analyses.

Its appealing properties notwithstanding, I am not advocating exclusive use of the exponentially decreasing volume density prior. My main message is rather that one should think carefully about what is an appropriate prior, given the available information, and what impact this has on the results.

\section*{Acknowledgements}

I would like to thank Morgan Fouesneau, Hans-Walter Rix, Anthony Brown, and Jan Rybizki, as well as two anonymous referees, for useful discussions and comments. This work was supported in part by the SFB 881 project ``The Milky Way System'' of the German Research Foundation (DFG).

\section*{References}

Bailer-Jones C.A.L., Andrae R., Arcay B. et al., 2013, {\em The Gaia astrophysical parameters inference system (Apsis). Pre-launch description}, A\&A 559, A74

Brown A.G.A, 1997, Arenou F., van Leeuwen F., Lindegren L., Luri X., 1997, {\em Considerations in making full use of the Hipparcos catalogue}, 	Proceedings of the ESA Symposium `Hipparcos - Venice '97', ESA SP-402 (July 1997), pp. 63--68

Brown A.G.A., 2012, {\em Statistical astrometry}, in ``Astrometry for Astrophysics'', van Altena W.F.\ (ed.), CUP

de Bruijne J.H.J., 2012, {\em Science performance of Gaia, ESA's space-astrometry mission}, Ap\&SS 341, 31--41

de Bruijne J.H.J., Rygl K.L.J., Antoja T., 2015, {\em Gaia astrometric science performance -- post-launch predictions}, arXiv:1502.00791

Francis C., 2013, {\em Calibration of RAVE distances to a large sample of Hipparcos stars}, MNRAS 436, 1343--1361

Ivezi{\'c} {\v Z}. and the LSST Science Collaboration, 2011, {\em Large Synoptic Survey Telescope (LSST) Science requirements document},  http://www.lsst.org/files/docs/SRD.pdf

Lindegren L., Babusiaux C., Bailer-Jones C. A. L., et al. 2008, {\em The Gaia mission: science, organization and present status}, in IAU Symposium vol. 248, ed.\ W.J.\ Jin, I.\ Platais, \& M. A. C. Perryman, 217–-223

Lutz T.E., Kelker D.H., 1973, {\em On the Use of Trigonometric Parallaxes for the Calibration of Luminosity Systems: Theory}, PASP 85, 573

Palmer M., Arenou F., Luri X., Masana E., 2014, {\em An updated maximum likelihood approach to open cluster distance determination},  A\&A 564, A49

Perryman M.A.C., Lindegren L., Kovalevsky J., et al., 1997, {\em The Hipparcos catalogue}, A\&A 323, L49

Robin A.C., Luri X., Reyl\'e C. et al., 2012, {\em Gaia Universe model snapshot. A statistical analysis of the expected contents of the Gaia catalogue}, A\&A 543, A100

Smith H., Eichhorn H., 1996, {\em On the estimation of distances from trigonometric parallaxes}, MNRAS 281, 211--218

Smith H., 2003, {\em Is there really a Lutz-Kelker bias? Reconsidering calibration with trigonometric parallaxes}, MNRAS 338, 891--902

Verbiest J. P. W., Weisberg J. M., Chael A. A., Lee K. J., Lorimer D. R., 2012, {\em On pulsar distance measurements and their uncertainties}, ApJ 755, 39

\end{document}